\documentclass[twocol]{ametsocV6.1}

\title{A new paradigm for medium-range severe weather forecasts: probabilistic random forest-based predictions}

\authors{Aaron J. Hill\aff{a}\correspondingauthor{Aaron J. Hill, aaron.hill@colostate.edu}, Russ S. Schumacher\aff{a}, and Israel Jirak\aff{b}}

\affiliation{\aff{a}{Colorado State University, Fort Collins, CO}\\
\aff{b}{Storm Prediction Center, NOAA}}

\abstract{Historical observations of severe weather and simulated severe weather environments (i.e., features) from the Global Ensemble Forecast System v12 (GEFSv12) Reforecast Dataset (GEFS/R) are used in conjunction to train and test random forest (RF) machine learning (ML) models to probabilistically forecast severe weather out to days 4--8. RFs are trained with $\sim$9 years of the GEFS/R and severe weather reports to establish statistical relationships. Feature engineering is briefly explored to examine alternative methods for gathering features around observed events, including simplifying features using spatial averaging and increasing the GEFS/R ensemble size with time-lagging. Validated RF models are tested with $\sim$1.5 years of real-time forecast output from the operational GEFSv12 ensemble and are evaluated alongside expert human-generated outlooks from the Storm Prediction Center (SPC). Both RF-based forecasts and SPC outlooks are skillful with respect to climatology at days 4 and 5 with degrading skill thereafter. The RF-based forecasts exhibit tendencies to underforecast severe weather events, but they tend to be well-calibrated at lower probability thresholds. Spatially averaging predictors during RF training allows for prior-day thermodynamic and kinematic environments to generate skillful forecasts, while time-lagging acts to expand the forecast areas, increasing resolution but decreasing objective skill. The results highlight the utility of ML-generated products to aid SPC forecast operations into the medium range.}

\begin{document}

%% Necessary!
\maketitle

%%%%%%%%%%%%%%%%%%%%%%%%%%%%%%%%%%%%%%%%%%%%%%%%%%%%%%%%%%%%%%%%%%%%%
% SIGNIFICANCE STATEMENT/CAPSULE SUMMARY
%%%%%%%%%%%%%%%%%%%%%%%%%%%%%%%%%%%%%%%%%%%%%%%%%%%%%%%%%%%%%%%%%%%%%
%
% If you are including an optional significance statement for a journal article or a required capsule summary for BAMS 
% (see www.ametsoc.org/ams/index.cfm/publications/authors/journal-and-bams-authors/formatting-and-manuscript-components for details), 
% please apply the necessary command as shown below:
%
% Significance Statement (all journals except BAMS)
%
\statement
%	 Enter significance statement here, no more than 120 words. See \url{www.ametsoc.org/index.cfm/ams/publications/author-information/significance-statements/} for details.

Medium-range forecasts generated from statistical models are explored here alongside operational forecasts from the Storm Prediction Center (SPC). Human forecasters at the SPC rely on traditional numerical weather prediction model output to make medium-range outlooks and statistical products that mimic operational forecasts can be used as tools for forecasters. The statistical models relate simulated severe weather environments from a global weather model to historical records of severe weather and perform noticeably better than human-generated outlooks at shorter lead times (e.g., day 4 and 5) and are capable of capturing the general location of severe weather events 8 days in advance. The results highlight the value in these data-driven methods in supporting operational forecasting.

%
%% Capsule (BAMS only)
%%
%\capsule
%       Enter BAMS capsule here, no more than 30 words. See \url{www.ametsoc.org/index.cfm/ams/publications/author-information/formatting-and-manuscript-components/#capsule} for details.
%
%% * * If using twocol mode, you will need to use the commands "twocolsig" and "twocolcapsule" in place of "sig" and "capsule"
%%      to ensure that the text box correctly spans across both columns.
%

%%%%%%%%%%%%%%%%%%%%%%%%%%%%%%%%%%%%%%%%%%%%%%%%%%%%%%%%%%%%%%%%%%%%%
% MAIN BODY OF PAPER
%%%%%%%%%%%%%%%%%%%%%%%%%%%%%%%%%%%%%%%%%%%%%%%%%%%%%%%%%%%%%%%%%%%%%
%

%% In all cases, if there is only one entry of this type within
%% the higher level heading, use the star form: 
%%
% \section{Section title}
% \subsection*{subsection one}
% % text...
% \section{Section title}

%vs

% \section{Section title}
% \subsection{subsection one}
% text...
% \subsection{subsection two}
% \section{Section title}

%%%
% \section{First primary heading}

% \subsection{First secondary heading}

% \subsubsection{First tertiary heading}

% \paragraph{First quaternary heading}

\section{Introduction}

Operational predictions of severe weather hazards (i.e., tornadoes, hail, and wind) are under the purview of the National Oceanic and Atmospheric Administration (NOAA) Storm Prediction Center (SPC), which is responsible for ``timely and accurate forecasts and watches for severe thunderstorms and tornadoes over the contiguous United States'' \citep{spcaboutus}. The SPC uses forecast guidance from numerical weather prediction (NWP) models, including post-processed products, diagnostic parameters (e.g., storm relative helicity), as well as current observations, to issue outlooks 1--8 days in advance of the threat of severe weather; outlooks issued in the 1 and 2-day timeframe delineate threats for specific hazards whereas day 3--8 outlooks highlight risk areas of any severe hazard. Because of the tremendous societal and financial impacts of severe weather events, including 10 severe-weather attributed billion-dollar disasters in 2021 alone \citep{NCEI2022}, it is imperative that SPC forecasters receive reliable and valuable forecast information to inform their operational products and provide sufficient lead time to stakeholders of the threat of severe weather.

Deterministic and ensemble NWP model predictions of severe weather and associated hazards have improved substantially over the last decade as dynamical models have leveraged increased computing power to decrease grid spacing and increase effective resolution. Increases in model resolution have seemingly benefited short-term forecasts the most, as real-time high-resolution models are now capable of explicitly resolving parent convective storms \citep[e.g.,][]{Done2004,Kain2008}; these prediction systems are commonly referred to as convection-allowing models (CAMs). These advances provided opportunities to explicitly forecast weather hazards (e.g., tornadoes) by using proxies \citep[e.g., updraft helicity;][]{Sobashetal2011,Sobashetal2016,Sobashetal2016b,Hilletal2021} in NWP model output, or generating calibrated guidance products \citep[e.g.,][]{Galloetal2018,Harrison2022,Jahnetal2022} that probabilistically depict severe weather threats like tornadoes and lightning. However, few of these methods have carried over into longer-range forecasting because of their dependence on CAMs, which are limited to the near-term ($<$ 4 days) due to computational constraints and undesirable, rapid increases in forecast spread from small-scale errors \citep{Zhangetal2003,Zhangetal2007}. As a result, SPC forecasters must rely on global prediction systems to issue day 4--8 outlooks (hereafter referred to as the medium range), including both deterministic and ensemble systems that parameterize convective processes and as a result, they have limited value for severe weather forecasting. 

Efforts have been made to leverage the large global ensemble datasets to generate post-processed and calibrated forecast products for medium-range forecast events as well. Post-processed fields become increasingly valuable at these extended lead times, offering a simplistic depiction for the threat of severe weather \citep[e.g., calibrated severe thunderstorm probabilities;][]{BrightandGrams2009} that may not be contained in highly-variable deterministic model output fields or ensemble output diagnostics (e.g., ensemble mean, variance). The U.S. National Blend of Models \citep{Hamilletal2017} uses quantile mapping to generate post-processed precipitation forecasts in the medium-range. Additionally, analog forecasting \citep{Lorenz1969} has been used to forecast high-impact weather environments \citep[e.g.,][]{HamillandWhitaker2006}, and typically involves training a regression model (e.g., logistic regression) on past environmental states that were coincident with severe weather reports and learning how similar environments relate to severe weather frequency (e.g., CIPS Analog-Based Severe Guidance\footnote{Cooperative Institute for Precipitating Systems (CIPS) Analog-Based Probability of Severe Guidance available at https://www.eas.slu.edu/CIPS/SVRprob/SPG\_Guidance\_whitepaper.pdf}). Using these developed statistical relationships, the current atmospheric patterns are interrogated to determine comparable analogs to base a new forecast. To the authors' knowledge, no analog techniques have been published addressing severe weather forecasting in the medium range, and a need still exists for techniques and tools to generate skillful medium-range severe weather forecast guidance.   

% [\citep{Melhauser2012} highlight the need for ensemble guidance rather than purely determinstic to improve severe weather predictability. 

More recently, more advanced machine learning (ML) models have emerged into the meteorology domain as a complementary and alternative option to forecast severe weather hazards. Whereas dynamical models cannot explicitly forecast hazards below their effective resolution, ML models can be trained to forecast any hazard given a sufficiently accurate observational dataset (i.e., labels) and related environmental features (i.e., predictors). ML has become a widely used technique to post-process NWP model output, for example, to generate forecasts mimicking operational center products \citep[e.g.,][]{herman2018money,Lokenetal2019,Hilletal2020,Lokenetal2020,Sobashetal2020,HillandSchumacher2021,Schumacheretal2021} and forecast severe weather or hazards more generally \citep[e.g.,][]{Gagne2014machine,gagne2017storm,Jergensenetal2020,McGovernetal2017,Burkeetal2020,Flora2021,Lokenetal2022}. Others have developed ML prediction systems using observations or reanalysis datasets \citep[e.g.,][]{Gensinietal2021,ShieldandHouston2022}, demonstrating ML-based prediction systems capable of highlighting conducive severe weather environments, for example. \citet{Hilletal2020} and \citet{Lokenetal2020} employed random forests \citep[RFs;][]{breiman2001random} to generate forecasts analogous to SPC outlooks, effectively creating post-processed and probabilistic first-guess forecasts of severe weather that could be used by forecasters when generating their human-based outlooks. \citet{Lokenetal2020} used CAM output to train an RF and derive day-1 hazard forecast probabilities whereas \citet{Hilletal2020} used global ensemble output to generate day 1-3 forecasts. Both studies demonstrated that RFs could produce skillful and reliable forecasts of severe weather at short lead times, and \citet{Hilletal2020} further highlighted that incorporating the statistical product information into the human-generated SPC forecast (i.e., through an statistical weighting procedure) yielded a better forecast at day 1 than either individual component forecast; the statistical models outperformed SPC forecasters at days 2 and 3. Therefore, it is reasonable to hypothesize that a similar prediction system devoted to forecasting severe hazards beyond day 3 would benefit SPC forecasters issuing medium-range forecasts. 

Building upon the work of \citet{Hilletal2020}, this study trains and develops RFs to forecast severe thunderstorm hazards in the medium range (i.e., days 4--8). The RF infrastructure of the Colorado State University Machine Learning Probabilities (CSU-MLP) prediction system \citep[e.g.,][]{Hilletal2020,Schumacheretal2021} is used herein to explore medium-range severe weather predictions and determine their utility in relation to operational forecasts. Feature engineering \citep[e.g.,][]{Lokenetal2022} is also explored (i.e., how to organize predictors from the dynamical model output used to train the RF models) to determine if medium-range ML-based forecast skill is impacted by the way predictors are gathered and trained with observations. All RF-based forecasts are evaluated alongside corresponding SPC outlooks to illustrate the relative value of incorporating the statistical guidance into forecast operations. 

\section{Methods}

\subsection{Random Forests}

RFs combine the simplicity of the decision tree architecture with the robustness of ensemble tree methods. Individual decision trees are constructed beginning with the root node (i.e., top of the tree) where a subset of training examples (i.e., instances of severe weather or no severe weather observation) is extracted from the full training set via bootstrapping and a feature is selected that best splits the examples, i.e., the feature best describes separation of severe weather events from non-events. Successive nodes are similarly constructed along the branches of the tree until a maximum tree depth is reached or a minimum number of training samples needed to split a node is breached, ending that particular branch in a terminal or “leaf” node. The leaf node either contains all the same observation types (e.g., all examples are severe weather events) or a mixture. As new inputs are supplied to a decision tree, e.g., from realtime forecast output, the tree can be traversed to a leaf node using the inputs, producing a deterministic or probabilistic prediction of severe weather from the single tree. The aggregation of all decision tree predictions from a forest of decision trees provides a probabilistic prediction for the threat of severe weather. 

\begin{figure*}[t]\centering
 \noindent\includegraphics[width=39pc,angle=0]{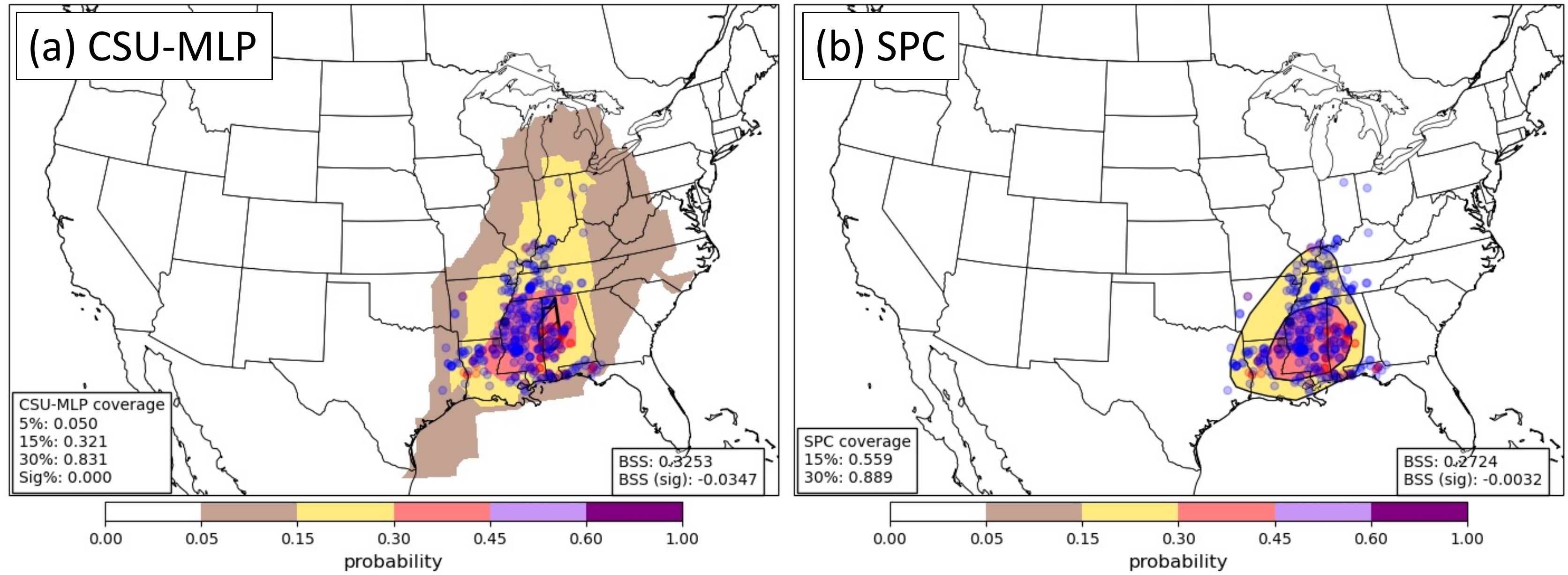}\\
 \caption{(a) CSU-MLP day 4 forecast initialized 27 March 2022, valid for period ending 1200 UTC 31 March 2022, and corresponding (b) day 4 SPC outlook. Forecast probabilities of any severe hazard are shaded, and the circle icons are SPC local storm reports of tornadoes (red), hail (green), and wind (blue). Forecast BSS and report coverage are provided in the lower right and left corners of each panel, respectively.}\label{EXAMPLE}
\end{figure*}

RF-based forecasts in this context are analogous to the SPC outlooks at days 4--8, i.e., the probability of severe weather occurrence within 40 km (or 25 mi) of a point over the daily 24-hour period defined by 1200--1200 UTC, and forecast products are constructed to mimic those produced by SPC (e.g., Fig. \ref{EXAMPLE}). For RF training, severe weather observations are encoded onto a grid by interpolating severe storm reports from the SPC Severe Weather Database \citep{spc2022} to NCEP grid 4 (0.5 degree grid spacing); the same grid is used to define RF features, further discussed in subsection 2.a.1. Any severe weather report during the 24-hour forecast window is interpolated to the 0.5 degree grid using a 40 km neighborhood, resulting in at least one grid point encoded for each severe weather report; the 0.5 degree grid has approximately 55 km grid spacing. Severe weather reports are defined as tornadoes of any EF rating, hail exceeding 1" (2.54 cm) in diameter, and convective wind speeds exceeding 58 mph (93 km/h). Training examples are encoded as 0 (no severe), 1 (non-significant severe), or 2 (significant severe) across the CONUS; the significant severe designation is a specific class of tornadoes eclipsing F2 or EF2 strength, hail exceeding 2” (5.08 cm) in diameter, and convective wind gusts exceeding 74 mph (119 km/h). Thus, the RFs are tasked with a prediction of 3 classes. However, because the current version of SPC day 4--8 outlooks do not include significant severe probability contours, CSU-MLP RF significant severe forecasts will not be formally evaluated in this work -- readers are referred to the work of \citet{Hilletal2020} who go in-depth on significant severe forecast skill for day-1 forecasts.

\subsubsection{Predictors}\label{subs1}

Meteorological features surrounding the encoded observations used for training are obtained from the Global Ensemble Forecast System version 12 (GEFSv12) Reforecast dataset \citep[hereafter GEFS/R;][]{Hamilletal2022,Zhouetal2022}, a 5-member daily 0000-UTC initialized ensemble system that utilizes the finite volume cubed sphere (FV3) dynamical core. Reforecasts date back to 1 January 2000 and extend forward to the end of 2019. Variables with known or presumed relationships to severe weather are extracted from the GEFS/R, including convective available potential energy (CAPE), precipitable water (PWAT), and bulk vertical wind shear (e.g., SHEAR500); a full list of variables is provided in Table \ref{t1}. Due to inconsistent resolution in the GEFS/R dataset, e.g., near-surface variables have higher resolution (0.25 degree grid) than upper-tropospheric variables (0.5 degree grid), each relevant meteorological output field is interpolated to the 0.5 degree grid, which also aligns the simulated reanalysis meteorological environments with the encoded observations. Additionally, latitude, longitude, and julian day are used as static features, coincident with the encoded severe weather report location. 

\begin{table*}[t]
\caption{Meteorological features used for RF training and forecasts}
\begin{center}
\begin{tabular}{ccc}
\hline\hline
Symbol & Variable Description & Variable Type\\
\hline
APCP & 3-hourly accumulated precipitation & thermodynamic \\
CAPE & Convective available potential energy & thermodynamic \\
CIN & Convective inhibition & thermodynamic \\
PWAT & Precipitable water & thermodynamic \\
T2M & 2-m temperature & thermodynamic \\
Q2M & 2-m specific humidity & thermodynamic \\
U10 & 10-m latitudinal horizontal wind speed & kinematic\\
V10 & 10-m longitudinal horizontal wind speed & kinematic\\
MSLP & Mean sea level pressure & kinematic\\
UV10 & 10-m wind speed & kinematic\\
SHEAR500 & Surface to 500 hPa bulk vertical wind difference & kinematic \\
SHEAR850 & Surface to 850 hPa bulk vertical wind difference & kinematic\\
\label{t1}
\end{tabular}
\end{center}
\end{table*}

\begin{figure*}[t]\centering
 \noindent\includegraphics[width=39pc,angle=0]{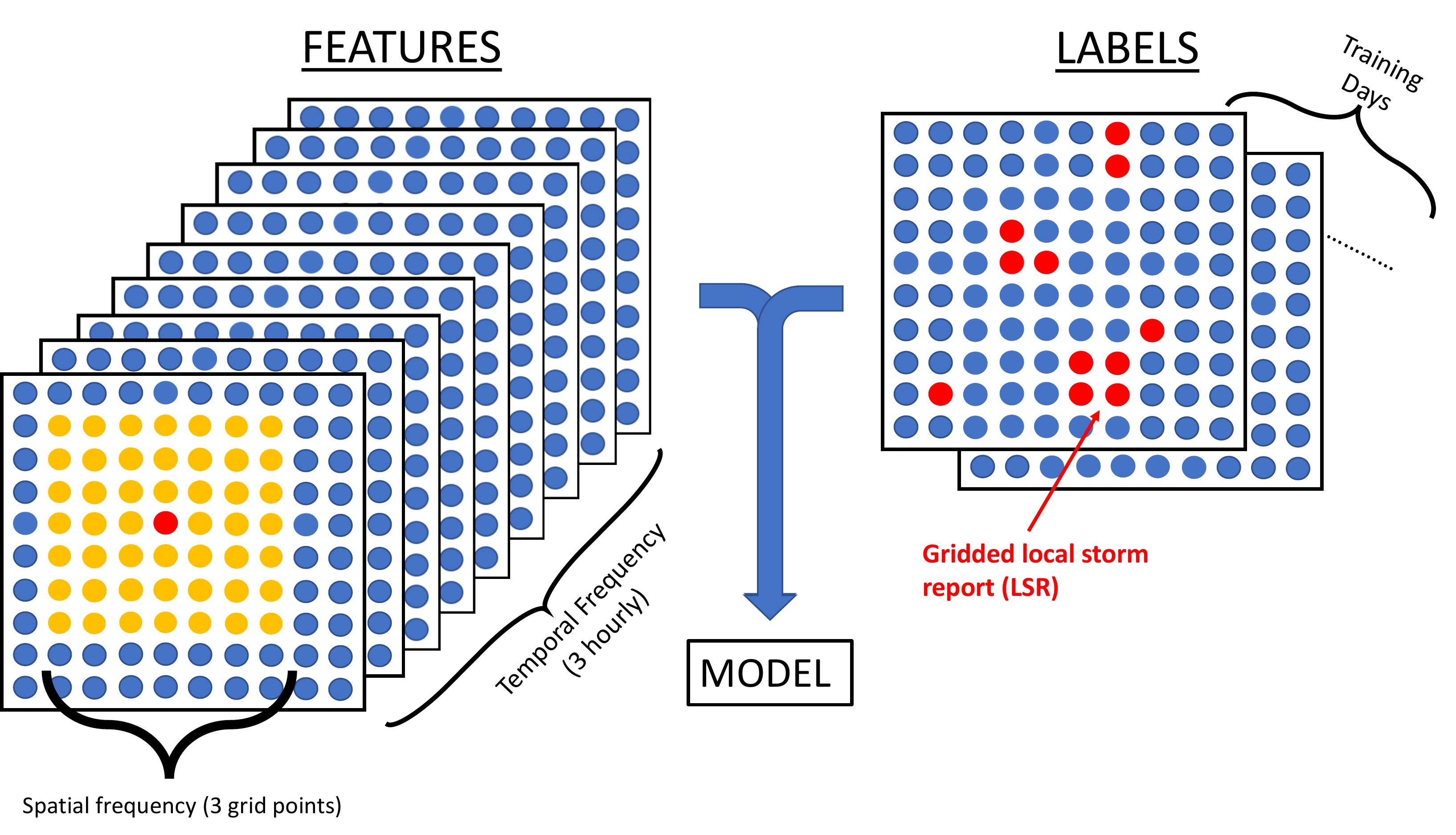}\\
 \caption{Schematic of feature and label assembly discussed in the text.}\label{ASSEMBLY}
\end{figure*}

Meteorological features are assembled in a forecast-point relative framework in which raw variables are gathered both at the observation training example point and over a pre-defined radius around the point. The pre-defined radius is set to 3 for all models trained in this manner. Additionally, the GEFS/R has 3-hourly temporal resolution across the 24 h forecast windows, allowing for both spatial and temporal sampling of the environment surrounding a severe weather report that occurred within the window; a depiction of the assembly method is provided in Fig. \ref{ASSEMBLY}. The raw variables accumulated in space and time represent the ensemble median for that given point; previous work demonstrated the superiority of using the ensemble median as opposed to the mean or outliers of the ensemble distribution \citep{herman2018money}. In other words, a 2-dimensional time series is constructed for each meteorological variable at every grid point across the forecast window.

\begin{figure*}[t]\centering
 \noindent\includegraphics[width=39pc,angle=0]{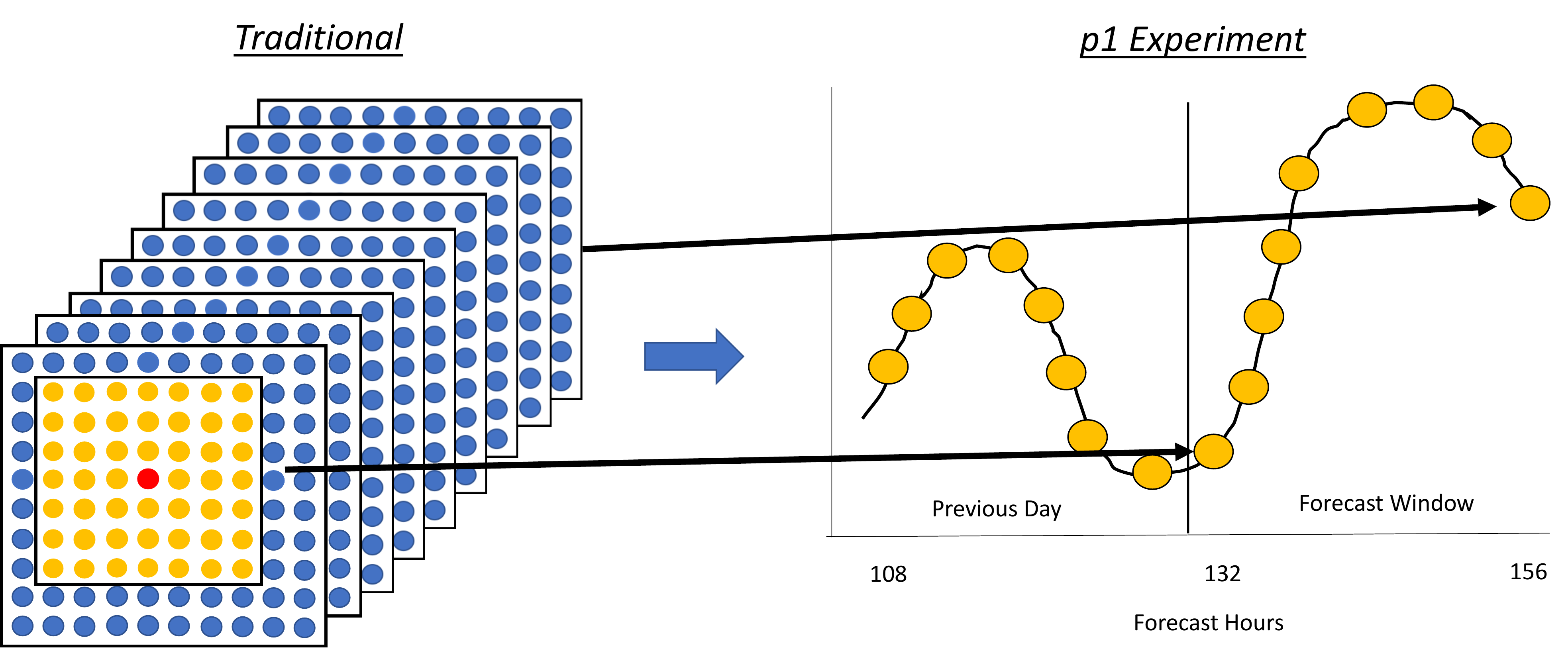}\\
 \caption{Schematic of feature assembly for the p1 model discussed in the text.}\label{ASSEMBLY_EXPERIMENT}
\end{figure*}

Two alternative feature assembly methods are also applied. Previous research employing the CSU-MLP prediction system has nearly universally applied the forecast-point relative framework to assemble features \citep[e.g.,][]{herman2018money,HermanandSchumacher2018b,Hilletal2020,Schumacheretal2021}. However, \citet{HillandSchumacher2021} showed that spatially averaging the features (e.g., 1-D time series) yielded improved RF-based excessive rainfall forecasts, and they hypothesized that characterizing the environment at a particular time with a single mean value reduced noise during model training. Similarly, \citet{Lokenetal2022} demonstrated that improved RF forecast skill could be achieved by using the ensemble mean at each spatial point rather than all individual members at each point because it reduced training noise. While \citet{HillandSchumacher2021} and  \citet{Lokenetal2022} both used CAM inputs to train RFs, which may have inherently more noise than a global model, the same spatial averaging procedure of \citet{HillandSchumacher2021} is employed here (e.g., Fig. \ref{ASSEMBLY_EXPERIMENT}) to explore if the medium-range RF predictions generated here suffer from noisy inputs and whether skill could be improved by removing that noise. 

Another unique opportunity presents itself when spatially-averaging predictors: the total number of features is significantly reduced, decreasing the computational time needed to train the RFs. For example, the original CSU-MLP feature assembly method (forecast-point relative method described above) contains $N=mp(2r + 1)^2$ features, where m is the number of meteorological variables ($m=12$), p is the number of forecast hours in the window ($p=9$), and r is the number of grid points used surrounding each training example ($r=3$); N is 5292 for the traditional CSU-MLP methods and 108 for the spatial-averaging procedure. The reduced number of features motivates an additional exploration into how preceding environments (i.e., the day(s) prior leading up to an event) may contribute to RF predictions and subsequent skill. For instance, it is well understood that moisture return from the Gulf of Mexico into the Great Plains can often precede a severe weather event, providing deeper boundary layer moisture and ample CAPE to support deep convection and severe weather \citep{JohnsandDoswell1992} -- can the RFs learn to better predict severe weather by considering how the atmosphere becomes `primed' before cyclogenesis in the lee of the Rockies? Furthermore, ensemble spread increases as a function of lead time, so considering GEFS/R ensemble median meteorological predictors on prior days for a day 6 forecast, for example, may yield additional confidence in a forecast outcome than considering only predictors spanning the day 6 forecast window. Therefore, three experiments are conducted that use the preceding 1, 2, and 3 days of features -- hereafter referred to as p1, p2, and p3, respectively. For simplicity, only the surrounding environment near the training example is sampled, rather than employing a trajectory analysis to sample the upstream environment; such a method could be explored in future work. Using four total days of features (i.e., 3 preceding days and 1 valid over the forecast window) only requires 396 features, an order of magnitude less than the traditional CSU-MLP method. For brevity, and since this method evaluation is exploratory in nature, the spatial averaging methodology, using 0--3 preceding days (i.e., p0, p1, p2, and p3), is only applied for day-6 model training and corresponding forecast evaluation. 

Time-lagging methods have also been shown to be effective at creating forecast spread, yielding improved forecasts over their deterministic components and are competitive with multi-model and initial-condition perturbation ensembles at convective scales \citep[e.g.,][]{Jiraketal2018,WadeandJirak2022}. One specific benefit from time-lagging is no additional computation is required since the forecasts are already created and can be combined relatively easily to compute ensemble statistics. In this work, time-lagging is used to artificially increase the GEFS/R ensemble size, ideally yielding a more representative ensemble median when assembling features. Both 10- and 15-member time-lagged ensembles are experimented with, which use the previous and 2 previous reforecast initializations, respectively, with features from each initialization valid over the same forecast window. As a result, the 10- and 15-member time-lagged RF models use 1 and 2 less days for training, discussed in the next subsection.

\subsubsection{Training, Validation, and Testing}

While the GEFS/R has 20 years of daily forecasts, only $\sim$9 years are using for training the medium-range forecast models. This decision was largely to facilitate a comparison between day 1--3 models developed by Hill et al. (2020) and companion models trained with the GEFS/R, which are not explored in this work. Daily initializations of the GEFS/R from 12 April 2003--11 April 2012 are used to assemble predictors and severe weather reports are aggregated and encoded for this same period. It should be noted that since the first initialization is 12 April 2003, the first training examples are valid for 1200--1200 UTC 15--16 April 2003. 4--fold cross validation is employed over the 9 year period to select an optimal model as well as avoid overfitting the RFs. Testing of the trained RFs is conducted 2 October 2020--1 April 2022; the operational GEFSv12 was implemented in early October 2020. 

\begin{figure*}[t]\centering
 \noindent\includegraphics[width=19pc,angle=0]{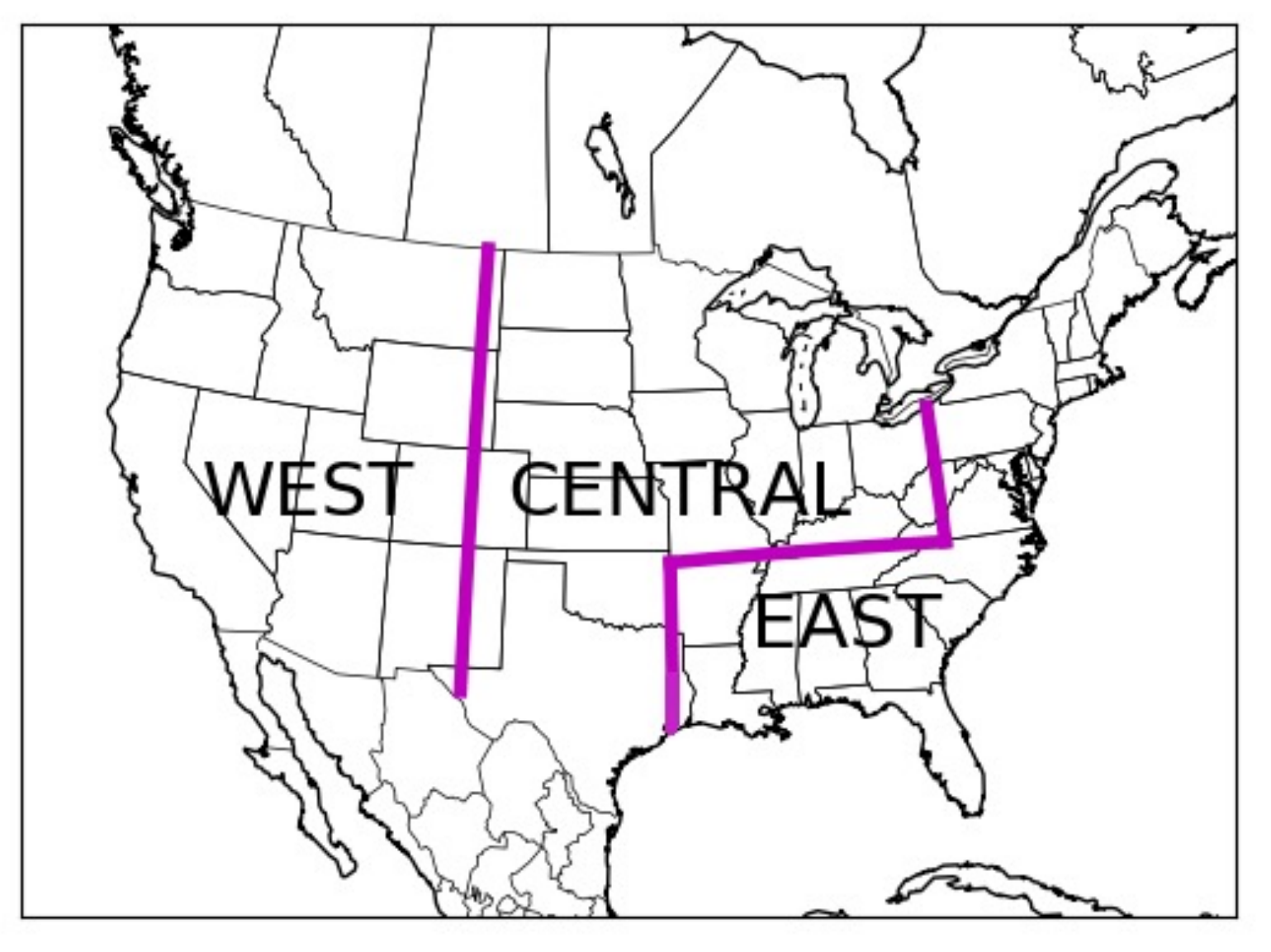}\\
 \caption{Region delineation for training the random forests. The west region is bounded by 20°–49°N, 240°–255°E; the central region by 25°–36.5°N, 255°–265.4°E and 36.5°–49°N, 255°–279.5°E; and the east region by 25°–29°N, 277°–280.2°E; 29°–36.5°N, 265.4°–285°E; and 36.5°–49°N, 279.5°–294°E}\label{REGION}
\end{figure*}

RFs are trained for distinct regions of the country that represent somewhat unique regional climatologies of severe weather (Fig. \ref{REGION}). For each region, an RF is cross validated and the optimal model is selected based on minimizing the Brier Score across the four folds. After a fold is chosen, hyperparameters are varied to tune the skillful RF; skill of the trained RFs is more sensitive to the cross validation fold than the hyperparameters varied. Hyperparameters varied included the number of trees in the forecast and minimum number of samples needed to split a node. Entropy was set as the splitting criterion and a random selection of the features were evaluated at each node, equal to the square root of the total number of features. RFs trained with the alternative predictor assembly methods undergo the same cross-validation and hyperparameter tuning procedure.  

Forecasts across the testing set are made using the GEFSv12 operational prediction system that consists of 21 ensemble members and 0.5 degree resolution. Forecasts are made with each regional, optimized RF and the severe weather probabilities are stitched together with a smoothing function to limit discontinuities at regional borders and create a CONUS-wide forecast. As with training, only the 21-member ensemble median is used to generate real-time features, which feed into all RF versions to generate predictions. It should be noted that the GEFS/R dataset could have been used for testing between 2012 and 2019, but realtime forecasts are generated with the operational GEFSv12 and the authors felt the most appropriate evaluation should use products that SPC forecasters would be using in the future. 

\subsection{Verification}

Traditional methods used to quantify probabilistic prediction skill -- e.g., Brier Skill Score (BSS), area under the receiver operating characteristic curve (AUROC), and reliability diagrams -- are employed to evaluate the CSU-MLP and SPC forecasts\footnote{SPC forecasters began looking at CSU-MLP forecasts in realtime beginning early 2022. As a result, the independence of RF forecasts and human-based outlooks cannot be guaranteed. Due to an already limited verification period, forecast dependence is ignored in the verification statistics.}. Observations of severe weather are obtained from the SPC archive of National Weather Service local storm reports since the SPC severe weather database was not updated through 2021 at the time the analysis was conducted. For a more direct comparison between continuous RF-generated forecasts and discrete SPC probabilities, the RF probabilities are discretized to resemble the outlooks issued by SPC forecasters. Discretization converts all RF probabilities within a probabilistic bin to the midpoint of the bin \citep[e.g.,][]{Schumacheretal2021} -- i.e., all probabilities below the 15\% minimum SPC probability contour are set to 7.5, probabilities between 15\% and 30\% to 22.5\%, and probabilities above 30\% are set to 65\%. The discretization procedure is also applied to the SPC contours. Additional discretized contours (e.g., a 5\% contour from the RF forecasts) are introduced in the verification section to elucidate factors influencing RF forecast skill. While the continuous RF probabilies could be evaluated alongside interpolated SPC contours \citep[e.g.,][]{Hilletal2020}, which have been shown to be more skillful than discrete contours \citep{Hermanetal2018c}, the limited number of possible SPC contours at days 4--8 reduces the utility of interpolation; the discrete 15\% and 30\% SPC contours are retained for verification. All medium-range SPC outlooks evaluated herein were issued at $\sim$0900 UTC each day and the shapefiles are converted to a gridded domain with ArcGIS as in \citet{Hermanetal2018c} as well as upscaled to NCEP Grid 4 for comparison to the RF-based forecasts.   

%,Karstens2019

The Brier Score (BS) is a measure of the mean squared error between binary events and probabilistic forecasts. The BS can be converted to a skill score, the BSS, by comparing the BS of a forecast to a reference climatology BS. BSS ranges from -$\infty$ to 1, with scores closer to 0 meaning the forecast skill is indistinguishable from the reference climatology ($BS_{ref}$):
\begin{equation}
    BSS = 1 - \frac{BS_{fcst}}{BS_{ref}}.
\end{equation}
The reference climatology used is a spatially and temporally smoothed long-term climatology of any severe weather reports across the CONUS. SPC severe weather database reports from 1990 to 2019 are gridded and Gaussian smoothers with 15-day temporal and 120-km spatial filters are used to create daily climatologies (e.g., Fig. \ref{CLIMO_FREQ}a) for the 30-year period \citep[e.g.,][]{Brooksetal2003,SobashandKain2017,Schumacheretal2021}. BSSs are computed in aggregate (i.e., considering all forecast points for all days) and spatially (i.e., considering all forecasts for a particular point in space) to characterize SPC and RF forecast skill. 

AUROC \citep[e.g.,][]{Marzban2004} measures forecast resolution and in this work the prediction system's ability to discriminate severe weather from non severe weather at various probabilistic thresholds. AUROC values of 0.5 suggest no discrimination, 0.7-0.8 is considered good, 0.8-0.9 great, and values $>$ 0.9 exceptional \citep{MANDREKAR2010}. Reliability diagrams are also used to characterize the relative forecast probability against the observed frequency of events, highlighting forecasts calibration at the various continuous probability contours and the discrete 15\% and 30\% SPC contours. Finally, the percent of forecast area covered by observations is computed to evaluate consistent biases in contour size. If a 15\% contour frequently contains more than 30\% fractional coverage of observations (i.e., the next contour interval), that issued contour is considered too small. Alternatively, if the fractional coverage is less than 15\%, the contour is too large. Fractional coverage has been used extensively to evaluate probabilistic ML-based forecasts against human-generated outlooks \citep[e.g.,][]{Ericksonetal2019,Hilletal2020,Ericksonetal2021,Hilletal2021,Schumacheretal2021} . 

Finally, the statistical relationships identified by the RFs are inspected to glean additional insights about the features that the models rely on to make predictions and how those relationships align with our physical understanding of severe weather forecasting. Feature importances (FIs) are computed and evaluated to assess the use of predictor information in each tree of the forest. Specifically, the "Gini importance" metric \citep[e.g.,][]{pedregosa2011scikit, HermanandSchumacher2018b, Whan2018, Hilletal2020} is used to quantify the FIs. Each feature is assigned an importance value based on the number of times it is used to split a decision tree node, weighted by the number of training examples at the split \citep{Friedman2001greedy}. The importances are then summed over all splits and trees in the forest and can be aggregated to characterize temporal or spatial importance patterns \citep[e.g.,][]{Hilletal2020}. The higher the importance, the more value the RFs place on that predictor to make predictions. While other FI techniques exist \citep[e.g.,][]{McGovern2019}, the Gini importance metric is used here as an initial glance and not a holistic interrogation of FIs of the developed RFs; a follow-on manuscript is being prepared to fully interrogate model FIs with sufficient breadth and depth.

\section{Verification Period Overview}
\subsection{Frequency of Severe Weather}

\begin{figure*}[t]\centering
 \noindent\includegraphics[width=39pc,angle=0]{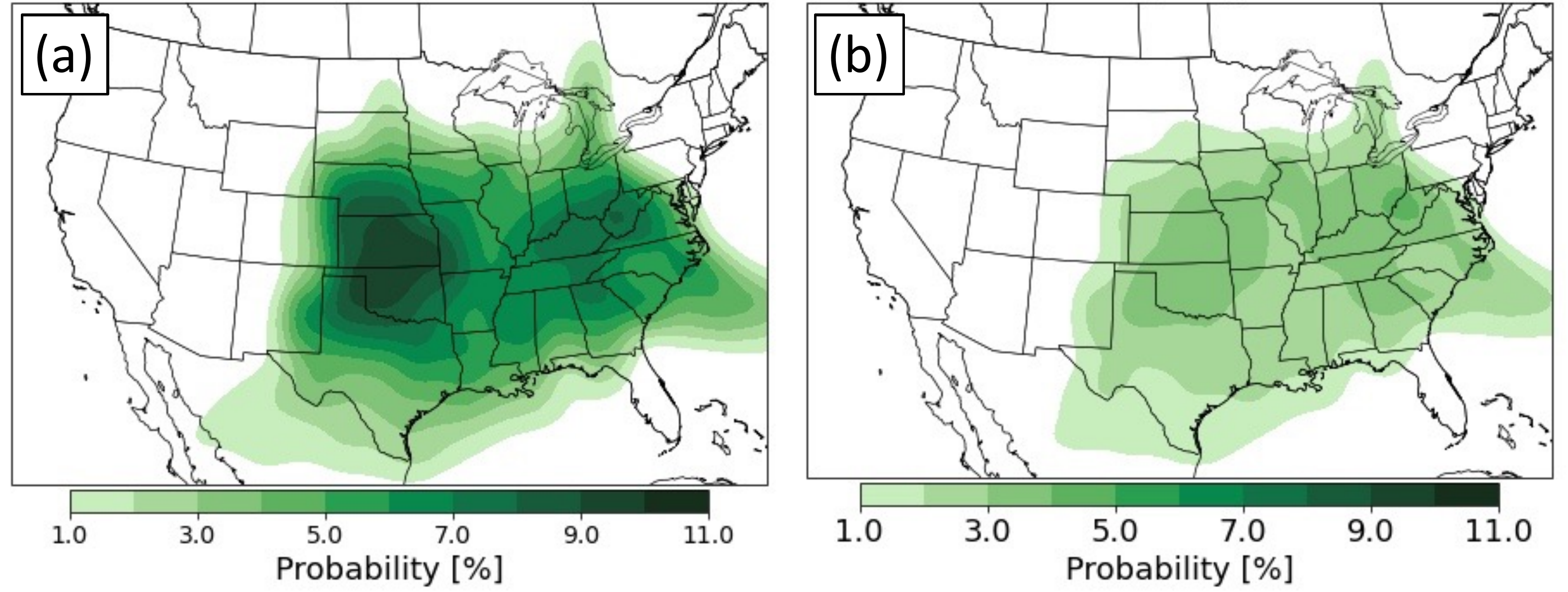}\\
 \caption{(a) Daily climatological any severe hazard probability (\%) centered on 21 May and (b) mean climatological hazard probability (\%) for any severe weather hazard.}\label{CLIMO_FREQ}
\end{figure*}

\begin{figure*}[t]\centering
 \noindent\includegraphics[width=39pc,angle=0]{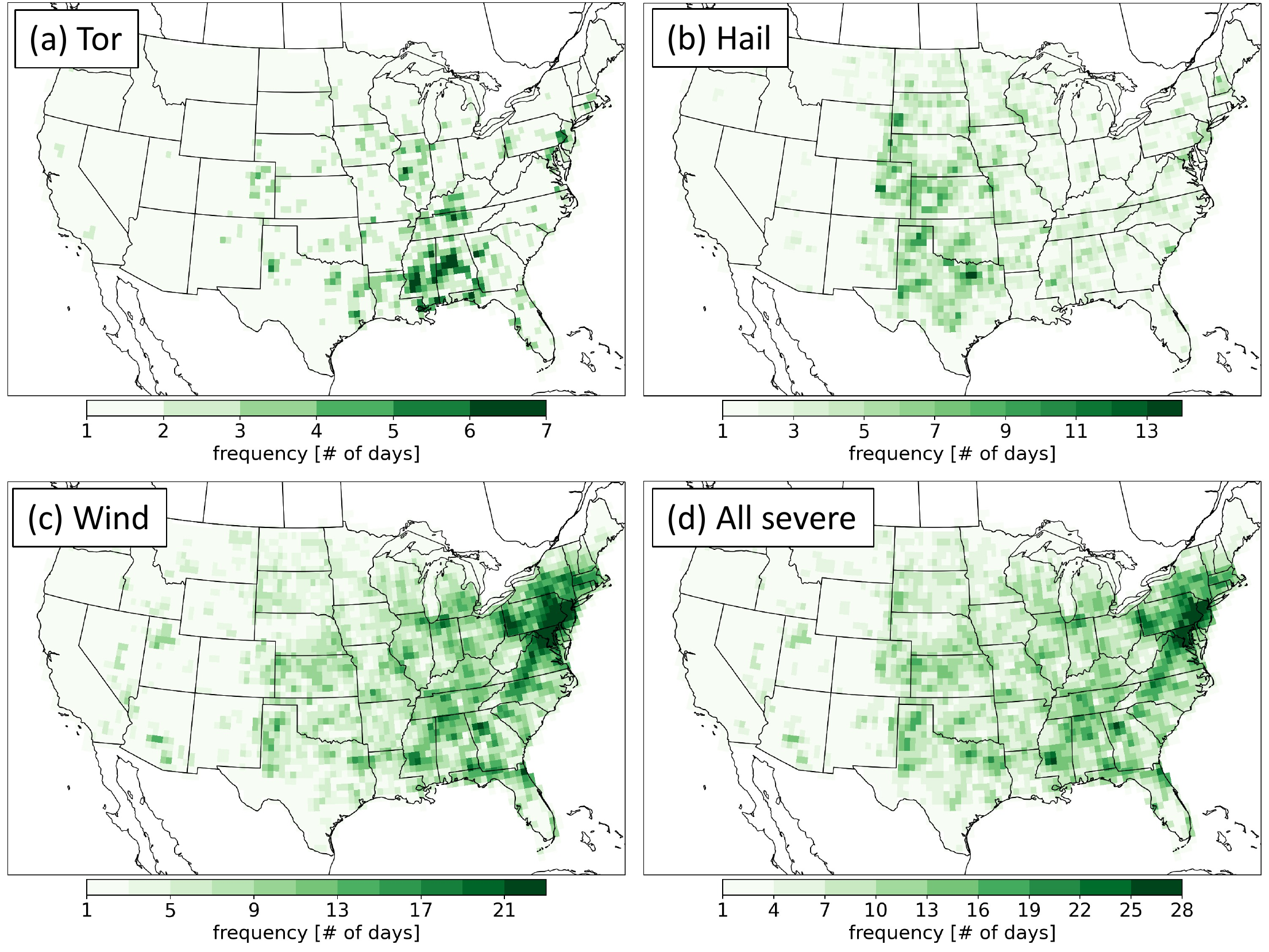}\\
 \caption{Frequency of (a) tornado, (b) hail, (c) wind, and (d) all severe hazard reports across the verification period.}\label{HAZ_FREQ}
\end{figure*}

Tornado event frequency highlights an active 1.5 years across the southeast U.S. (Fig. \ref{HAZ_FREQ}a), likely attributable to two fall-winter seasons in the dataset; the climatological frequency in tornado activity across the southeast US has two peaks, one in the early fall and the other late winter/early spring \citep[e.g.,][]{Horgan2007,Guyer2010,Dixon2011,GensiniandAshley2011,Cintineo2012hail,smith2012torclimo,Gensinietal2020}. With only one full spring season of severe weather in 2021, limited tornadoes were reported across the Great Plains (e.g., Fig. \ref{HAZ_FREQ}b). Unsurprisingly, hail reports are primarily confined to the Great Plains and wind reports are more uniform across the U.S. (Fig. \ref{HAZ_FREQ}c), east of the Rocky Mountains, compared to the other two hazards. Across the mid-Atlantic up into New England, multiple high-impact weather events produced numerous wind reports (Fig. \ref{HAZ_FREQ}c,d). These reports were anomalous compared to the long-term severe weather climatology (Fig. \ref{CLIMO_FREQ}b). 

% \subsection{SPC Outlooks}
% Arguably one of the most important aspects of ML applications in meteorology is comparing an ML-based forecast to a pre-existing baseline. For example, short-term ML-based forecasts of severe weather may be compared to simulated updraft helicity (UH) swaths from a high-resolution NWP model, calculated using various vertical layers (e.g, 0--3 km), as UH statistically relates to instances of severe weather hazards (Sobash paper) and has been used extensively in CAM verification studies (e.g., Hill et al. 2021). Alternatively, others have evaluated ML-based forecasts alongside human-based forecasts (e.g., Hill et al. 2020, Schumacher et al. 2021, Hill and Schumacher 2021, and others) that represent the same predictand as being predicted (e.g., severe weather report within 25 km of a point). Given the lack of simulated high-resolution and storm-specific variables (e.g., UH) at extended ranges, the RF-based forecasts discussed hereafter are directly compared to operational forecasts disseminated by expert SPC forecasters. 

\subsection{Frequency of Forecasts over extended period (day 4--8)}

\begin{figure*}[t]\centering
 \noindent\includegraphics[height=45pc,angle=0]{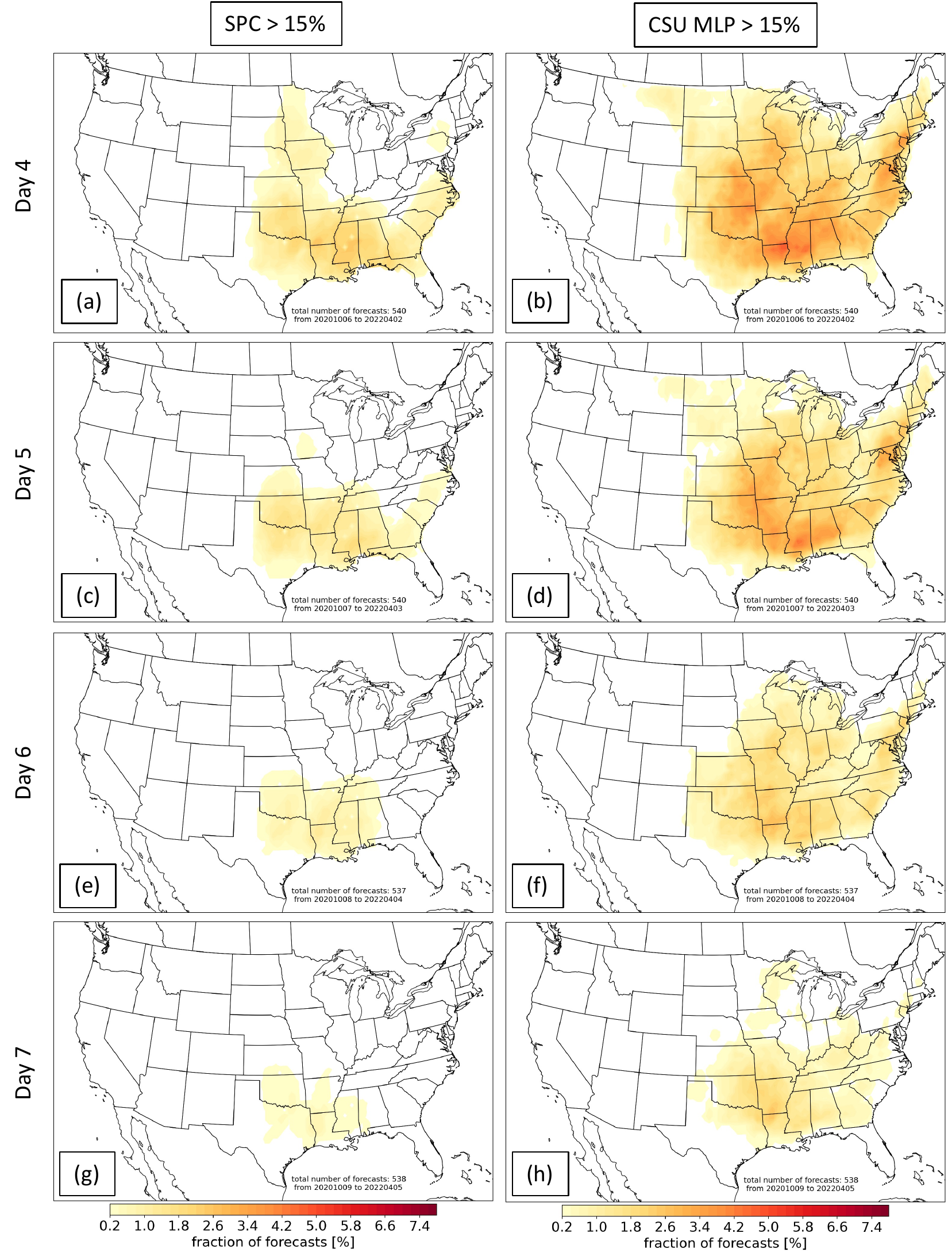}\\
 \caption{Fraction of day 4--7 (left) SPC and (right) CSU-MLP forecasts (\% of verification days) at at least the 15\% probability level. }\label{FCST_FREQ}
\end{figure*}

Day 4--7 15\% forecasts from the CSU-MLP system and outlooks from SPC were issued across areas that experienced frequent severe weather, including the southeast US and to some extent the southern Great Plains out to day 7 (Fig. \ref{FCST_FREQ}); 30\% forecast contours are qualitatively similar, but omitted for brevity. The RF-based forecasts were issued much more frequently, however, compared to SPC. At day 7, where SPC issued only a handful of 15\% contours across the CONUS (Fig. \ref{FCST_FREQ}g), the RF issued forecasts in some point locations as many as 10 times (2\% of the days; Fig. \ref{FCST_FREQ}h). Moreover, the RFs forecast areas of severe weather across larger areas of the CONUS, covering nearly all states east of the Rockies at day 5 (Fig. \ref{FCST_FREQ}d) when SPC limited their day 5 outlooks primarily south of 37$^o$ N (i.e., the northern border of Oklahoma). 

\begin{figure*}[t]\centering
 \noindent\includegraphics[width=39pc,angle=0]{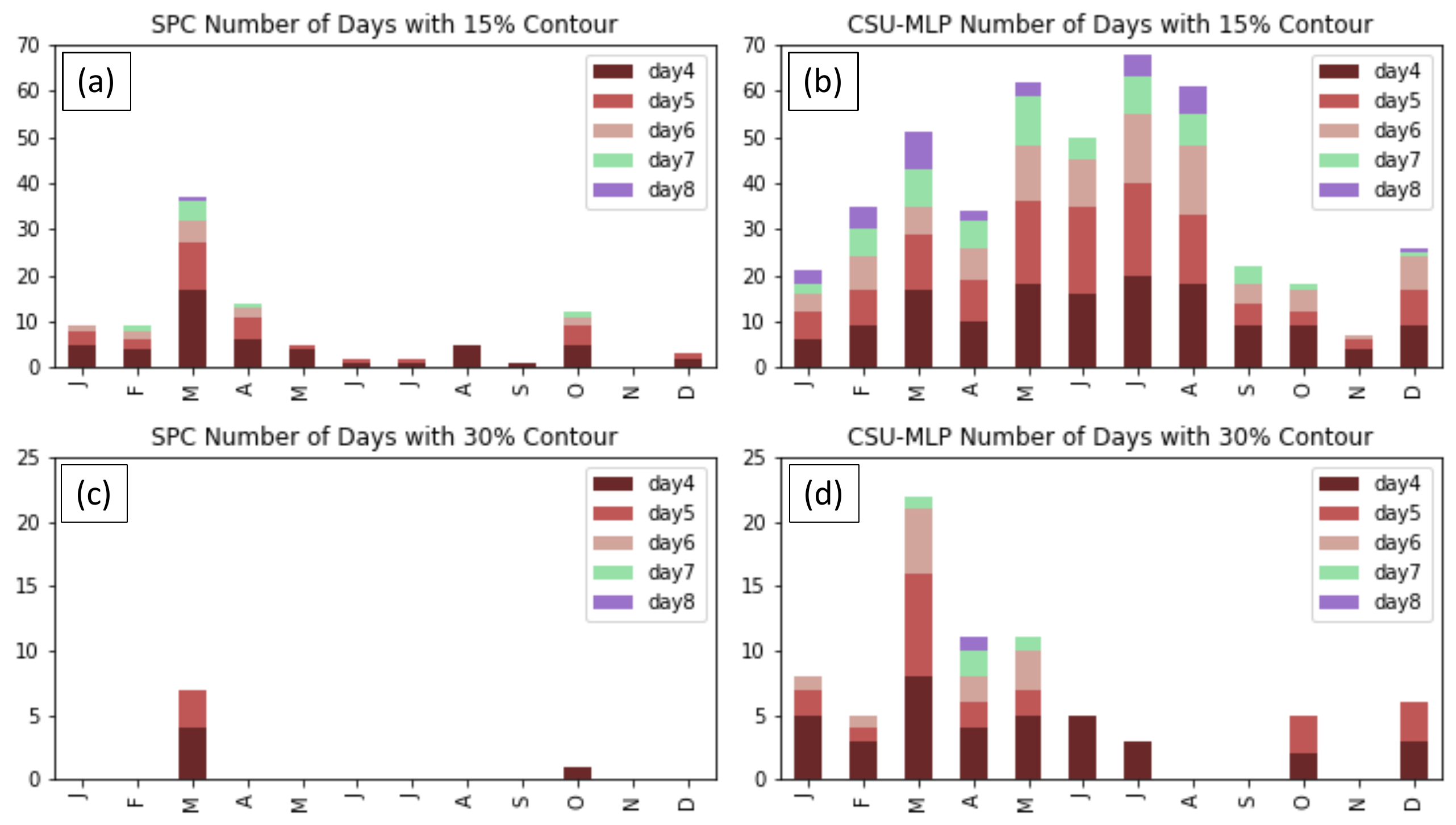}\\
 \caption{Frequency of day 4--8 (a),(c) SPC and (b),(d) CSU-MLP forecasts by month for (top) 15\% and (bottom) 30\% probabilistic contours. Forecast days are color coded per the legend and frequencies are stacked.}\label{MON_FREQ}
\end{figure*}

Despite the relatively short verification period, and only one spring forecast season in the verification dataset, there is clearly seasonality to the issuance of SPC medium-range outlooks and CSU-MLP forecasts. SPC issued more 15\% contours in the spring and fall seasons than summer and winter (Fig. \ref{MON_FREQ}a). This pattern is not necessarily surprising given the climatological "double peak" of severe weather across the CONUS in the spring and fall \citep[e.g.,][]{smith2012torclimo}. In contrast, the CSU-MLP had a nearly uniform distribution of 15\% contours across the months for days 4--6 (Fig. \ref{MON_FREQ}b), with perhaps a slight peak in frequency across the summer months -- 15\% contours at days 7 and 8 were more common between January and August. SPC issued only a handful of 30\% contours in the months of March and October (Fig. \ref{MON_FREQ}c) whereas the CSU-MLP issued a number of higher probability contours through the spring months, primarily at days 4 and 5 (Fig. \ref{MON_FREQ}d), highlighting the RF-based system's confidence in forecasting both predictable and less predictable (i.e., warm season) severe weather regimes.    

\section{Forecast Verification}

\begin{figure*}[t]\centering
 \noindent\includegraphics[width=39pc,angle=0]{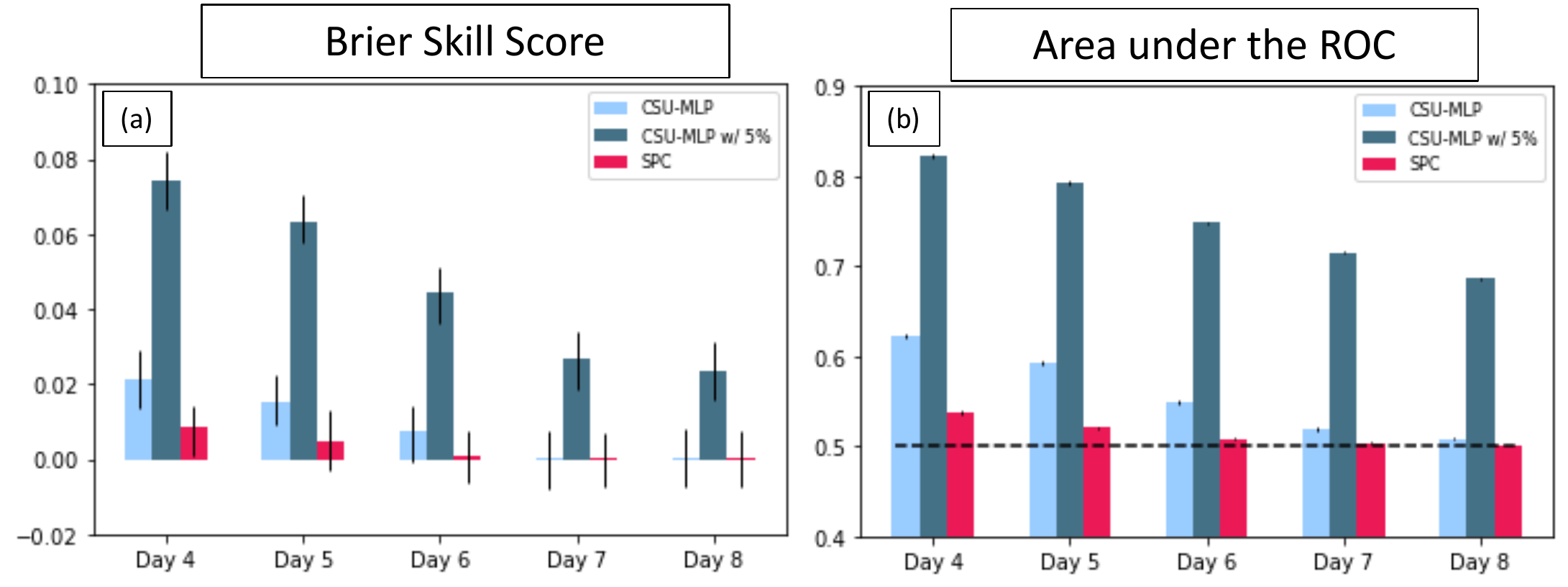}\\
 \caption{Aggregate (a) Brier Skill Score and (b) Area under the ROC (AUROC) curve for CSU-MLP RF forecasts and SPC outlooks. Included as dark blue is BSS and AUROC for CSU-MLP forecasts that include the 5\% contour. Error bars are computed from 100 bootstrap resamples of the forecast distributions and represent the 95\% confidence interval.}\label{SKILL}
\end{figure*}

RF and SPC forecast skill is first evaluated in aggregate across the entire verification period. Control RF (i.e., CSU-MLP) forecast skill decreases with increasing lead time, demonstrating significantly better skill than SPC outlooks at days 4 and 5 and equally near-climatological skill beyond day 5 (Fig. \ref{SKILL}a). As lead time increases, the RFs are confronted with learning how GEFS/R environments (specifically, the median environments) associate with severe weather as forecast variability and ensemble variance also increases. The limited number of GEFS/R ensemble members likely prohibits a proper depiction of all future atmospheric states and forces the RFs to "learn" relationships between severe weather events and simulated environments that may not be conducive to severe weather. As a result, the ability of the RFs to discriminate events from non-events similarly decreases with increasing lead time, with AUROC falling to $\sim$0.5 by day 8 (Fig. \ref{SKILL}b); the AUROC is highest for day 4 at 0.62. However, none of the AUROC values eclipse the 0.7 mark denoting good resolution \citep{MANDREKAR2010}. On the other hand, when 5\% probability contours are included in the RF-forecast discretization process and retained in the AUROC calculation, reliability is increased substantially to $>$0.07 at day 4 and is significantly larger than 0 (i.e., climatology) at day 8. Meanwhile, the AUROC surpasses 0.8 at day 4 and resolution is nearly `good' at day 8. While the higher probability 15 and 30\% contours do not have significant skill beyond day 5 (or good resolution at any forecast lead time), the adjusted skill metrics resulting from including 5\% probability contours demonstrates that low-probability forecast contours may have tremendous value for forecasting severe weather out to day 8. 

\begin{figure*}[t]\centering
 \noindent\includegraphics[width=39pc,angle=0]{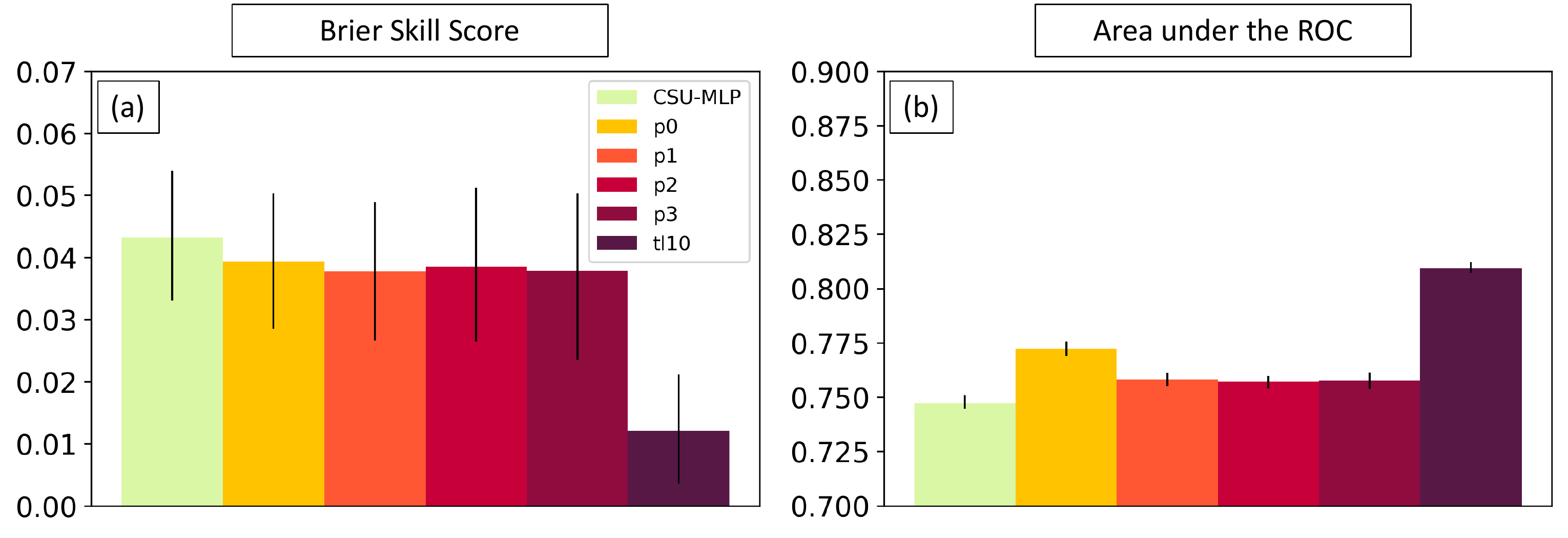}\\
 \caption{Aggregate (a) Brier Skill Score and (b) Area under the ROC (AUROC) curve for CSU-MLP and experimental (p0, p1, p2, p3, and tl10) RF forecasts. Error bars are computed from 100 bootstrap resamples of the forecast distributions and represent the 95\% confidence interval. BSS and AUROC are computed using 5, 15, and 30\% probability contours. }\label{SKILL_EXP}
\end{figure*}

The skill and resolution of forecasts derived from alternatively-trained RFs is also assessed to determine if medium-range prediction skill can be improved by learning to associate severe weather events with features in the days leading up to events (p0, p1, p2, and p3 experiments), reducing the impact of noisy predictors, and increasing the representative sample of atmospheric states in the underlying GEFS/R ensemble (tl10 experiment) used to assemble associated features. Aggregated BSSs are computed over the same verification period for the p0, p1, p2, p3, and tl10 experimental forecasts at the day 6 lead time (Fig. \ref{SKILL_EXP}a) with the 5\% contour included for comparison against the best control CSU-MLP forecasts (e.g., Fig. \ref{SKILL}a). The forecast skill for all predictor-averaged models is statistically indistinguishable from the control CSU-MLP RF model (Fig. \ref{SKILL_EXP}a). These BSSs alone imply that reasonably skillful day-6 forecasts can be derived by simply considering how the atmosphere evolves locally before a high-impact severe weather event. Moreover, the relatively equal skill amongst forecasts suggests that the raw GEFS/R predictors used in the control model (e.g., at each point in space) do not add significant value in training and perhaps exhibit less noise than their CAM-model counterparts \citep[e.g.,][]{Hilletal2021,Lokenetal2022}. The time-lagged model forecasts exhibit significantly less skill than the control and predictor-averaged model forecasts, but their resolution is significantly better (Fig. \ref{SKILL_EXP}b) suggesting that the time-lagged RF model is issuing probability contours that are larger than the corresponding predictor-averaged model forecast contours (i.e., improved resolution), but sacrificing skill (not shown). Furthermore, the predictor-averaged model forecasts all improve upon the resolution of the control system, with p0 containing the highest resolution (Fig. \ref{SKILL_EXP}b). 

%BSS over just days when SPC OR RF has a minimum (15\%) contour, or time series of skill?

\begin{figure*}[t]\centering
 \noindent\includegraphics[height=40pc,angle=0]{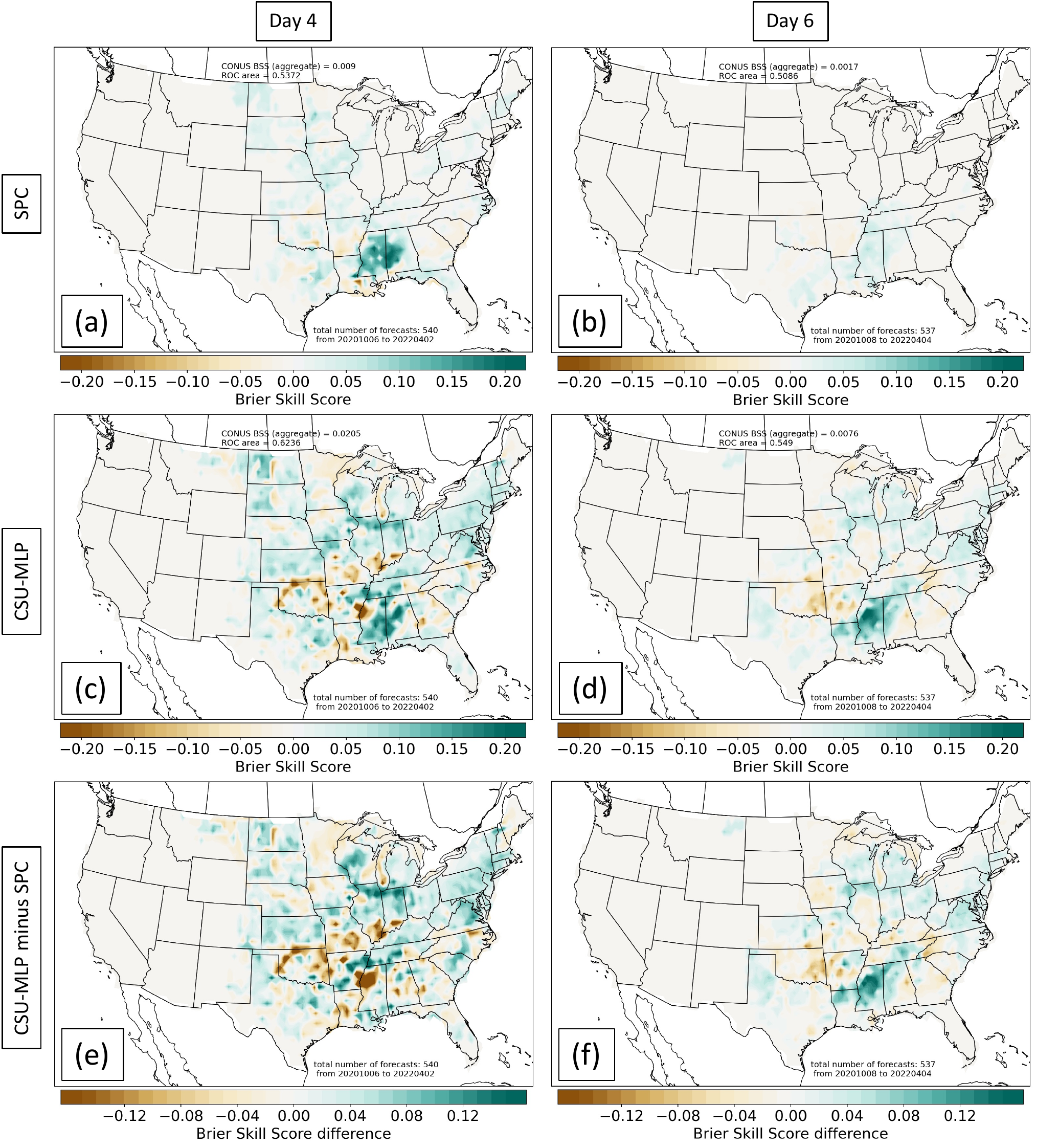}\\
 \caption{Aggregate BSS across the CONUS for (a),(c) day 4 and (b),(d) day 6 (top row) SPC and (middle row) CSU-MLP RF forecasts. CSU-MLP minus SPC BSS is shaded in the bottom row. Browns indicate where SPC forecasts are more skillful, and greens where CSU-MLP forecasts are more skillful.}\label{SKILL_SPACE}
\end{figure*}

An assessment of spatial forecast skill underscores the frequency biases of the RF-based and SPC forecasts. The SPC outlooks feature patchy areas of positive skill associated with instances of severe weather that were correctly highlighted days in advance (Figs. \ref{SKILL_SPACE}a,b); in other words, when SPC forecasters do issue outlooks, they do so quite skillfully, particularly at day 4. However, areas of slightly negative skill in the southern Great Plains and mid-Atlantic suggest there were missed opportunities to forecast high-impact events 4 and 6 days in advance. Fewer SPC forecasts at day 6 (Fig. \ref{FCST_FREQ}e) further limits the extent of positive forecast skill (Fig. \ref{SKILL_SPACE}b) compared to day 4. It is possible that forecasters had little confidence in the forecast evolution to warrant a 15\% or 30\% outlook contour at these lead times for various high-impact weather events, but as lead time decreased (i.e., days 1--3), confidence increased and forecasts were issued at multiple probabilistic thresholds. Furthermore, the limited verification dataset likely creates localized areas of high/low skill, and increasing the length of verification to multiple years would help to clarify SPC forecast skill in the medium range. 

\begin{figure*}[t]\centering
 \noindent\includegraphics[width=19pc,angle=0]{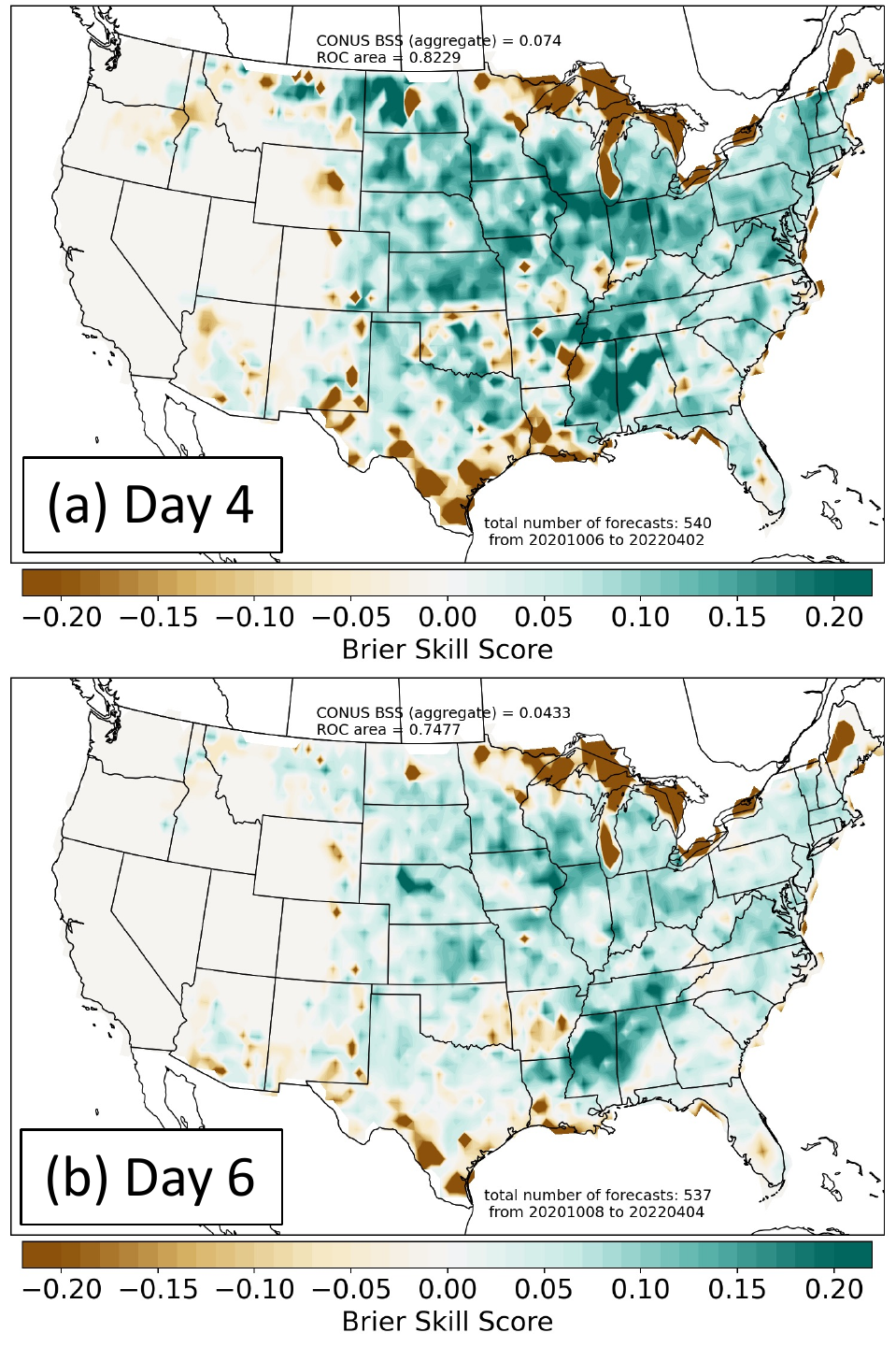}\\
 \caption{As in Figs. \ref{SKILL_SPACE}c,d, but BSS is computed with a discretized 5\% contour.}\label{SKILL_SPACE_INCLUDE}
\end{figure*}

The spatial skill of control RFs similarly emphasizes the high-frequency forecast bias with more expansive and smoothed areas of positive and negative BSSs (Fig. \ref{SKILL_SPACE}c,d). For brevity and since the predictor-averaged models display similar skill, only the control forecast spatial skill is considered here. Prominent areas of negative forecast skill in the control RF forecasts across the Great Lakes is a symptom of the "forecast anywhere" nature of the RFs regardless of the climatological report frequency. BSS differences between SPC outlooks and RF forecasts (Fig. \ref{SKILL_SPACE}e,f) further illustrate the complexities of verifying these forecasts on a short period, but also clearly demonstrate that issuing more probabilistic contours in the medium-range yields better skill. This facet is perhaps most notable across the upper Midwest and mid-Atlantic, where the RF forecasts at days 4 and 6 were notably better than SPC as a result of SPC not issuing many outlooks (Figs. \ref{FCST_FREQ}a,e). The CSU-MLP spatial forecast skill at days 4 and 6 is improved when the 5\% probability contour is included (Fig. \ref{SKILL_SPACE_INCLUDE}), which amplifies areas of positive and negative skill. In some instances, including the 5\% forecast contour reverses areas of negative skill (c.f. Figs. \ref{SKILL_SPACE}d and \ref{SKILL_SPACE_INCLUDE}b in Georgia) or replaces neutral skill with positive skill as in the Great Plains (c.f. Figs. \ref{SKILL_SPACE}d and \ref{SKILL_SPACE_INCLUDE}b in North and South Dakota).  

\begin{figure*}[t]\centering
 \noindent\includegraphics[width=39pc,angle=0]{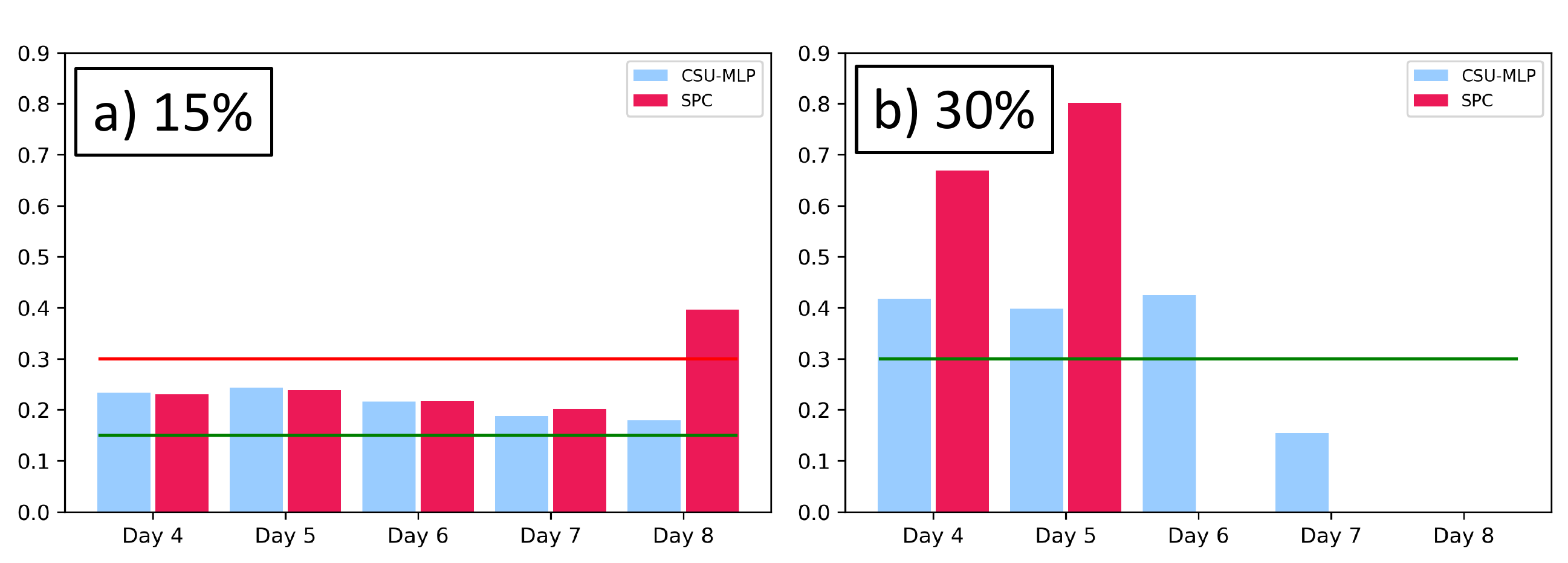}\\
 \caption{Spatial coverage of local storm reports in SPC and CSU-MLP forecasts for all forecast days and the (a) 15\% and (b) 30\% probability contours. Green horizontal lines denote the bottom probability of the forecast "bin", whereas red horizontal lines denote the top of the bin, i.e., the next probability contour value.}\label{COVERAGE}
\end{figure*}

Forecast skill is also assessed by computing the fractional coverage of observations within the forecast contours. While the SPC contours are defined as single probabilistic levels and not a discrete range (e.g., 15--30\%), it is reasonable to suggest that the fractional coverage of observations for a probabilistic contour should not exceed the next probabilistic value, otherwise a higher contour would be warranted. Fig. \ref{COVERAGE} shows the fractional coverage of observations, in which the objective is to be within the green and red horizontal bars for a particular probabilistic threshold. When below the green horizontal line, a forecast contours are believed to be too large on average, and when above the red line, forecast areas are too small. SPC and CSU-MLP control forecasts are well calibrated at the 15\% threshold at almost all forecast days; SPC forecasts are potentially too small at day 8, but a small sample size limits a complete analysis for that lead time (Fig. \ref{COVERAGE}a). On the other hand, the day 4 and 5 30\% outlooks from SPC are typically smaller than the CSU-MLP control forecasts, which appear generally well-calibrated prior to day 7 (Fig. \ref{COVERAGE}b). Unsurprisingly, the CSU-MLP 5\% probability contours are also well-calibrated at all forecast lead times (not shown).

\begin{figure*}[t]\centering
 \noindent\includegraphics[width=39pc,angle=0]{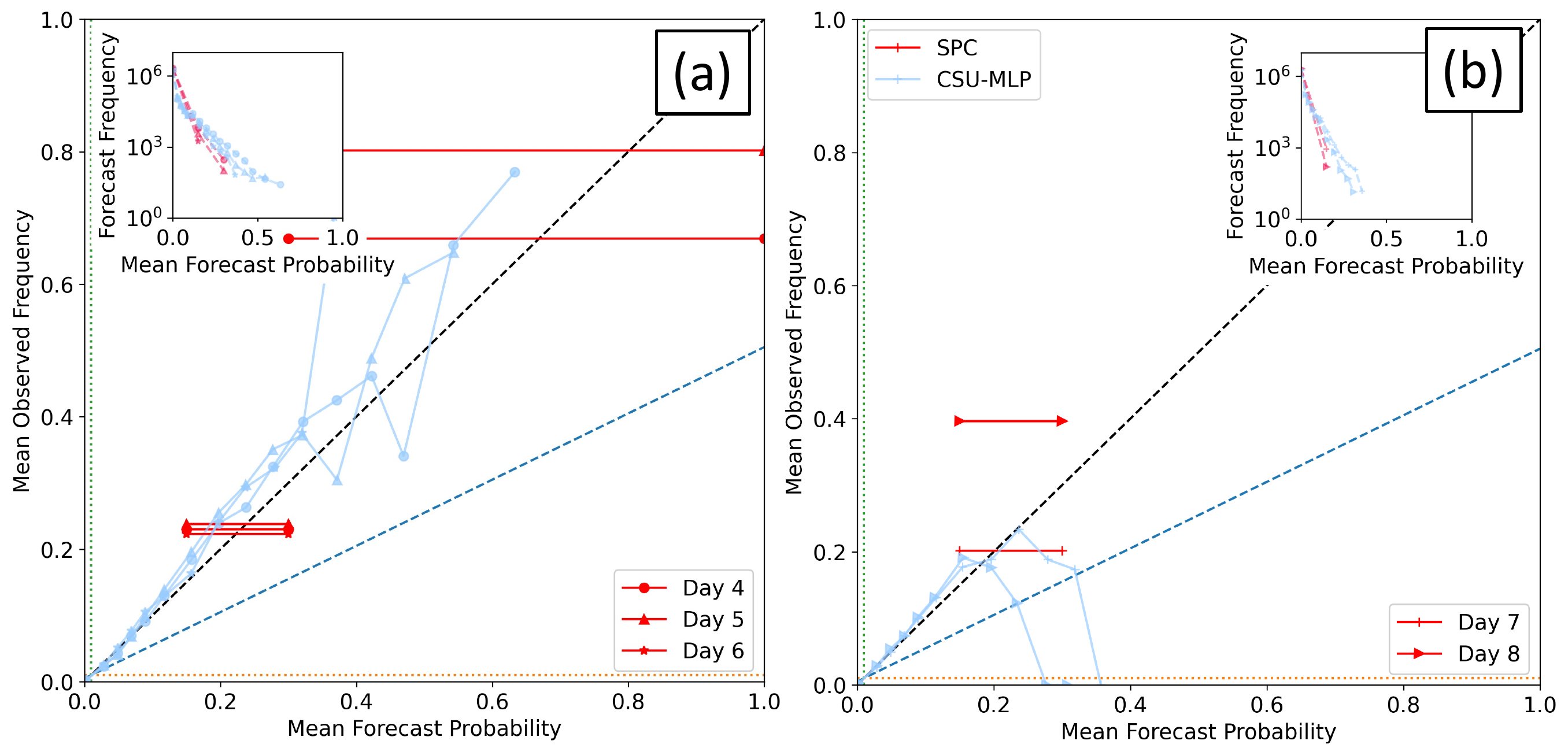}\\
 \caption{Reliability diagrams for CSU-MLP RF and SPC outlooks at (a) day 4--6 and (b) day 7--8 lead times and as noted in the legends. SPC reliability is plotted as the observed frequency across the forecast probability bins (e.g., 15-30\%). Inset are the forecast frequencies as a function of mean forecast probabilities. Perfect reliability, no skill, and no resolution are denoted by the black, blue, and green (or orange) dashed lines, respectively. }\label{REL_DIAG}
\end{figure*}

For a complete depiction of calibration, reliability diagrams are constructed for all SPC and control RF forecasts (Fig. \ref{REL_DIAG}). Reliability curves that fall above or below the perfect-reliability line, dashed black lines in both panels of Fig. \ref{REL_DIAG}, are said to under-forecast or over-forecast severe weather events, respectively. All SPC outlooks are generally well-calibrated prior to day 8, whereas RF forecasts generally have an under-forecast bias above the 15\% probability threshold but achieve reliability for lower thresholds. Day 7 and 8 control RF forecasts lose reliability quickly after 15\% (Fig. \ref{REL_DIAG}b), plummeting effectively to no skill above 30\% and no resolution (i.e., below the red dashed line) at the highest probabilities considered -- SPC maintains skill at day 7 but underforecasts severe weather events at day 8 (Fig. \ref{REL_DIAG}b). While reliability for days 4--6 forecasts above 30\% is considerably more variable (Fig. \ref{REL_DIAG}a), owing to smaller sample sizes (e.g., inset figure in Fig. \ref{REL_DIAG}), forecast skill still hovers near perfect reliability. Analysis of reliability for the alternative predictor assembly methods reveals that predictor-averaging maintains reliability relative to the control forecasts up to slightly higher probabilistic thresholds (e.g., 25\%) for the day 6 forecasts whereas the day 6 tl10 experiments overforecast observed events (not shown), aligning with the aggregate BSS statistics (Fig. \ref{SKILL_EXP}a). This analysis underscores the utility of the continuous RF probabilities as a forecast tool at medium-range lead times, and also stresses the difficulty of accurately capturing severe weather threat areas multiple days in advance (i.e., most RF-based forecasts underforecast the observed event probabilities), even for skillful statistical models. 

% \begin{figure*}[t]\centering
%  \noindent\includegraphics[width=19pc,angle=0]{reliability_exps.pdf}\\
%  \caption{Reliability diagrams for day 6 RF forecasts from the CSU-MLP control and experimental systems as denoted in the figure legend. Perfect reliability, no skill, and no resolution are denoted by the black, blue, and green (or orange) dashed lines, respectively. }\label{REL_DIAG_EXP}
% \end{figure*}

\subsection{Example Forecasts}

\begin{figure*}[t]\centering
 \noindent\includegraphics[width=39pc,angle=0]{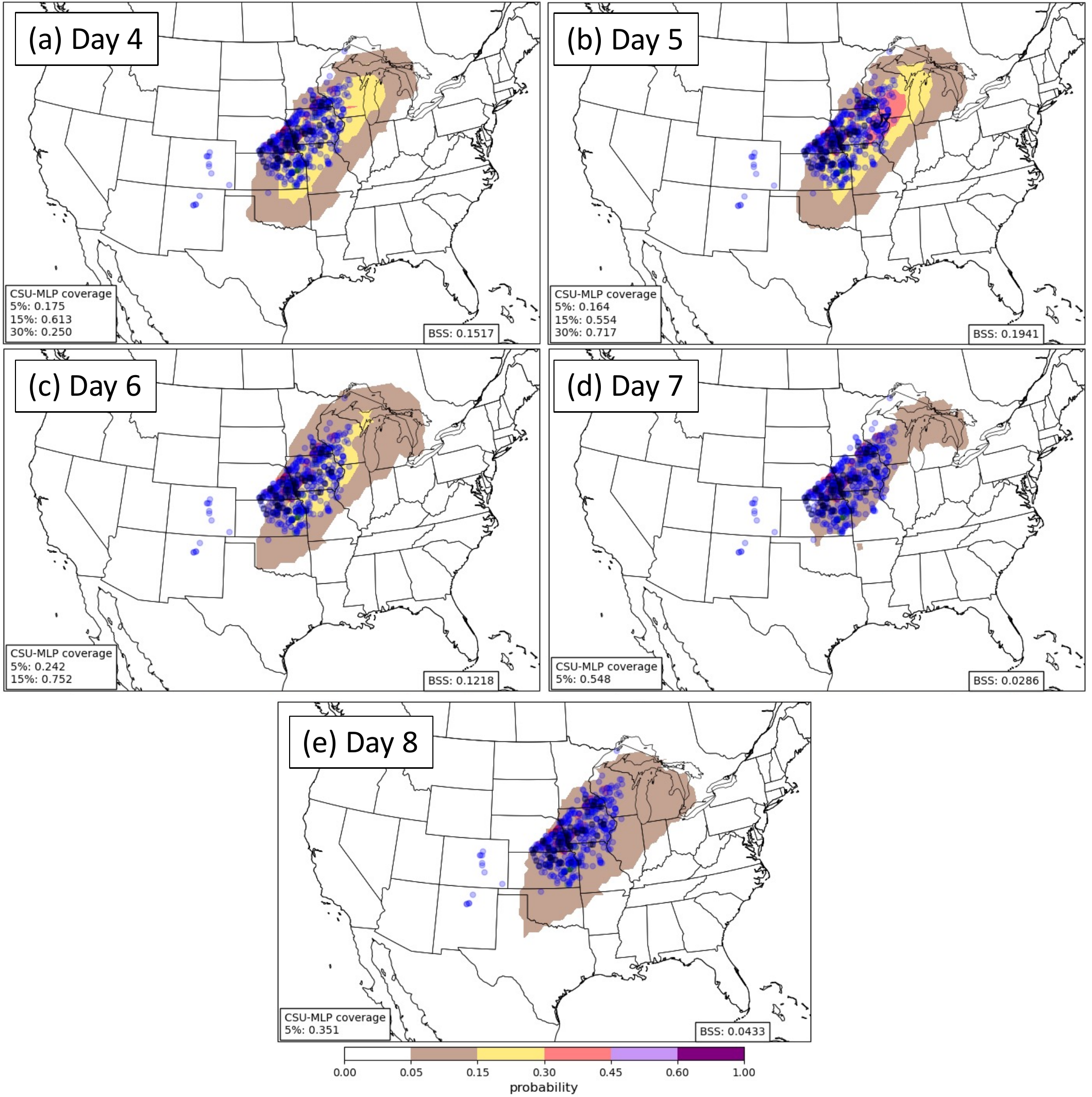}\\
 \caption{Day 4--8 CSU-MLP forecasts for any severe hazard probability valid 1200 UTC 15 December 2021 -- 1200 UTC 16 December 2021. NWS local storm reports for wind, hail, and tornadoes are included as blue, green, and red circles, respectively. Observation coverage and BSS are included in lower left and right corners of each panel, respectively.}\label{GOOD_CASE}
\end{figure*}

Example forecasts are provided that display some of the skill and resolution attributes of the RF outlooks described previously (Fig. \ref{GOOD_CASE}). All day 4--8 example forecasts are valid for the 24-hour period ending 1200 UTC 16 December 2021. This particular event featured a compact short-wave trough with strong low- and mid-level wind fields. Robust low-level warm air advection with low 60s dewpoints contributed to an atmosphere primed for severe thunderstorms and a highly anomalous severe-weather event for mid-December; a robust discussion of the meteorological parameter space for this event is provided in the day-1 SPC forecast discussion archive \citep{SPC2022b}. Medium-range forecasts from NWP models depicted the shortwave trough days in advance, but did not have a good handle on the instability parameter space even at day 4. SPC forecasters decided to issue a 5\% severe hazard risk at day 3, noting the impressive kinematic support for damaging wind gusts\footnote{Day 3 discussion available at https://www.spc.noaa.gov/products/outlook/archive/2021/day3otlk\_20211213\_0830.html}. Increasing forecaster confidence in a high-impact severe weather event with decreasing lead time resulted in a moderate categorical risk being issued at day 1, with a 45\% probability of severe wind and a large area of significant severe wind delineated; a 10\% probability of tornadoes also accompanied the SPC day-1 forecast. 

The CSU-MLP control forecasts (Fig. \ref{GOOD_CASE}) depicted a severe weather threat area across the upper Mississippi valley eight days in advance (Fig. \ref{GOOD_CASE}e). By day 6, a 15\% probability contour was introduced in the forecasts with a 30\% contour added in subsequent, shorter lead-time forecasts (Fig. \ref{GOOD_CASE}a-c) that mostly encircled severe weather reports for the event. Forecast skill scores (BSS) gradually increased from 0.04 at day 8 to 0.19 at day 5 with a slight decrease back to 0.15 at day 4. In short, the probabilistic RF-based guidance showed substantial skill out to day 8 for this particular case, and the progression of forecasts from day 8 to 4 showcases the utility of the forecast system to highlight areas that may experience severe weather\footnote{SPC forecasters used the CSU-MLP forecasts during this event, highlighting their value in upgrading SPC outlooks as the event neared. See note from SPC forecaster Andrew Lyons: \url{https://twitter.com/TwisterKidMedia/status/1471585397440487433?s=20&t=cSYwf08xjtvvuwIYr2TgHQ}}. 

% \begin{figure*}[t]\centering
%  \noindent\includegraphics[width=39pc,angle=0]{033022_spc.pdf}\\
%  \caption{Day 4--7 SPC outlooks for any severe hazard probability valid 1200 UTC 30 March 2022 -- 1200 UTC 31 March 2022. NWS local storm reports for wind, hail, and tornadoes are included as blue, green, and red circles, respectively. Observation coverage and BSS are included in lower left and right corners of each panel, respectively.}\label{GOOD_CASE_SPC}
% \end{figure*}

%\subsection{Alternative Assembly Methods}

\begin{figure*}[t]\centering
 \noindent\includegraphics[width=39pc,angle=0]{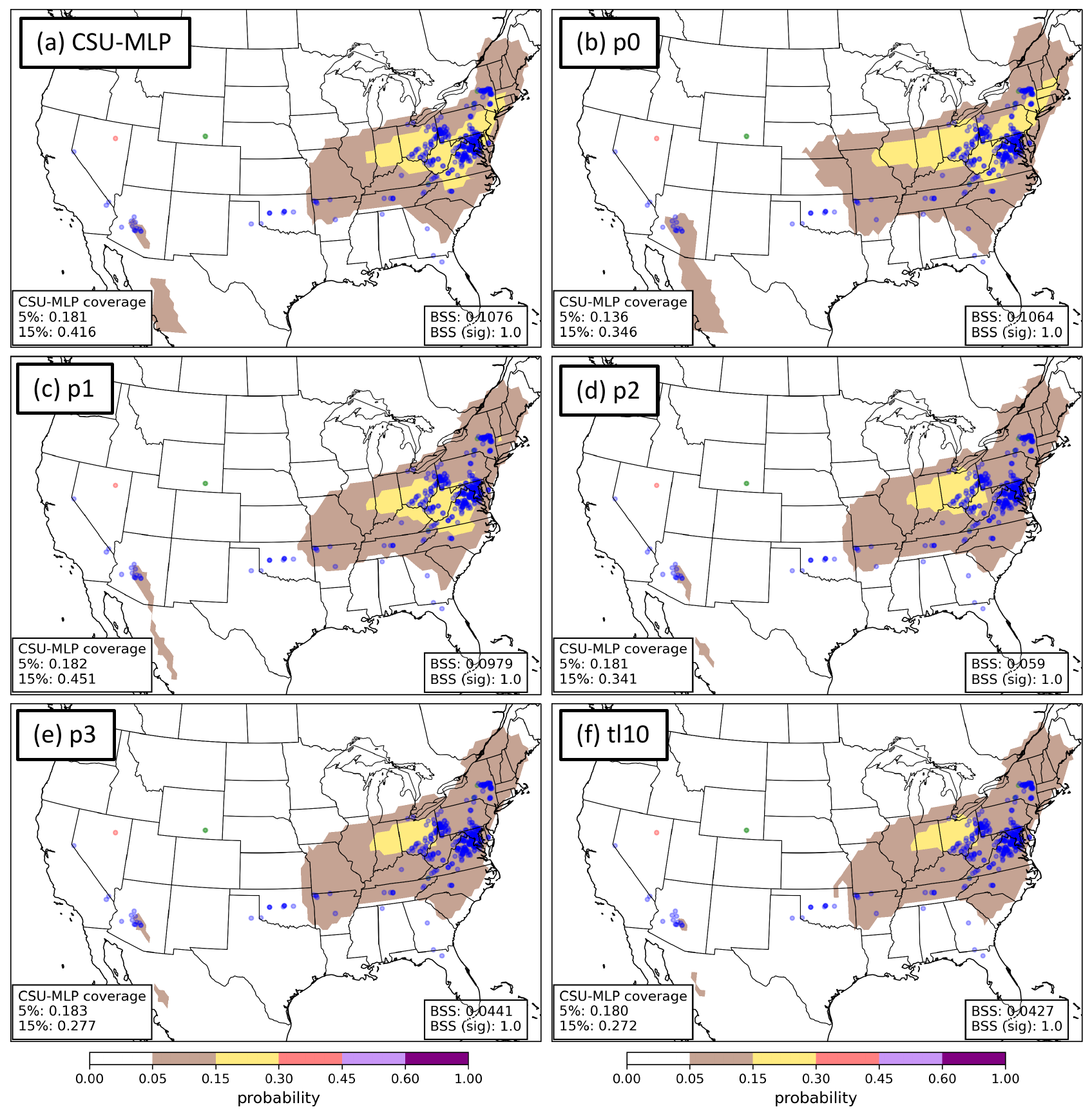}\\
 \caption{RF-based forecasts of any severe hazard at day 6 lead time from (a) trad, (b) p0, (c) p1, (d) p2, (e) p3, and (f) tl10. All forecasts are initialized 0000 UTC 8 August 2021 and valid 1200 UTC 13 August 2021 -- 1200 UTC 14 August 2021. NWS local storm reports for wind, hail, and tornadoes are included as blue, green, and red circles, respectively. Observation coverage and BSS are included in lower left and right corners of each panel, respectively.}\label{EXP_EX}
\end{figure*}

Another example is provided to reinforce the similarities between the experimental forecast systems for one particular case. On this day, the 24-hour period ending 1200 UTC 14 August 2021, numerous wind reports were recorded across Ohio, Pennsylvania, West Virginia, and numerous other mid-Atlantic states (Fig. \ref{EXP_EX}). All RF-based forecasts have a broad 5\% contour across this region, and erroneously westward into Indiana, Illinois, and Missouri. None of the forecasts suggest greater than 15\% probability of severe weather, despite dense sets of wind reports in two corridors across the northeast. The 15\% contour in the control system seems subjectively well-positioned (BSS=0.1076), but the p0-model forecast inaccurately extends the 15\% contour westward (BSS=0.1064). The p1-model forecast contracts the 15\% contour back eastward, but also eliminates an area in New York and Pennsylvania that experience wind reports (BSS falls to 0.0979). The p2- and p3-model forecasts further contract the 15\% contour (BSSs of 0.059 and 0.0441, respectively), leaving the 5\% area relatively unchanged; the time-lagged model is nearly identical to the p3-model forecast. This example illustrates the rather subtle differences between the experiments that renders the control system and flow-dependent RF models objectively similar. A more comprehensive case-study evaluation would be needed to characterize these subtle forecast differences over the entire forecast period, which is beyond the scope of this experimental exploration but is an active area of research. Additionally, the computational savings of the spatial-averaging models, particularly in training the RFs, may support a continued investigation alongside the CSU-MLP control forecast system into their utility as an operational tool. 

% \begin{figure*}[t]\centering
%  \noindent\includegraphics[width=32pc,angle=0]{122821.pdf}\\
%  \caption{28 December 2021 bust}\label{BAD_CASE}
% \end{figure*}

\subsection{Feature Importances}

\begin{figure*}[t]\centering
 \noindent\includegraphics[width=39pc,angle=0]{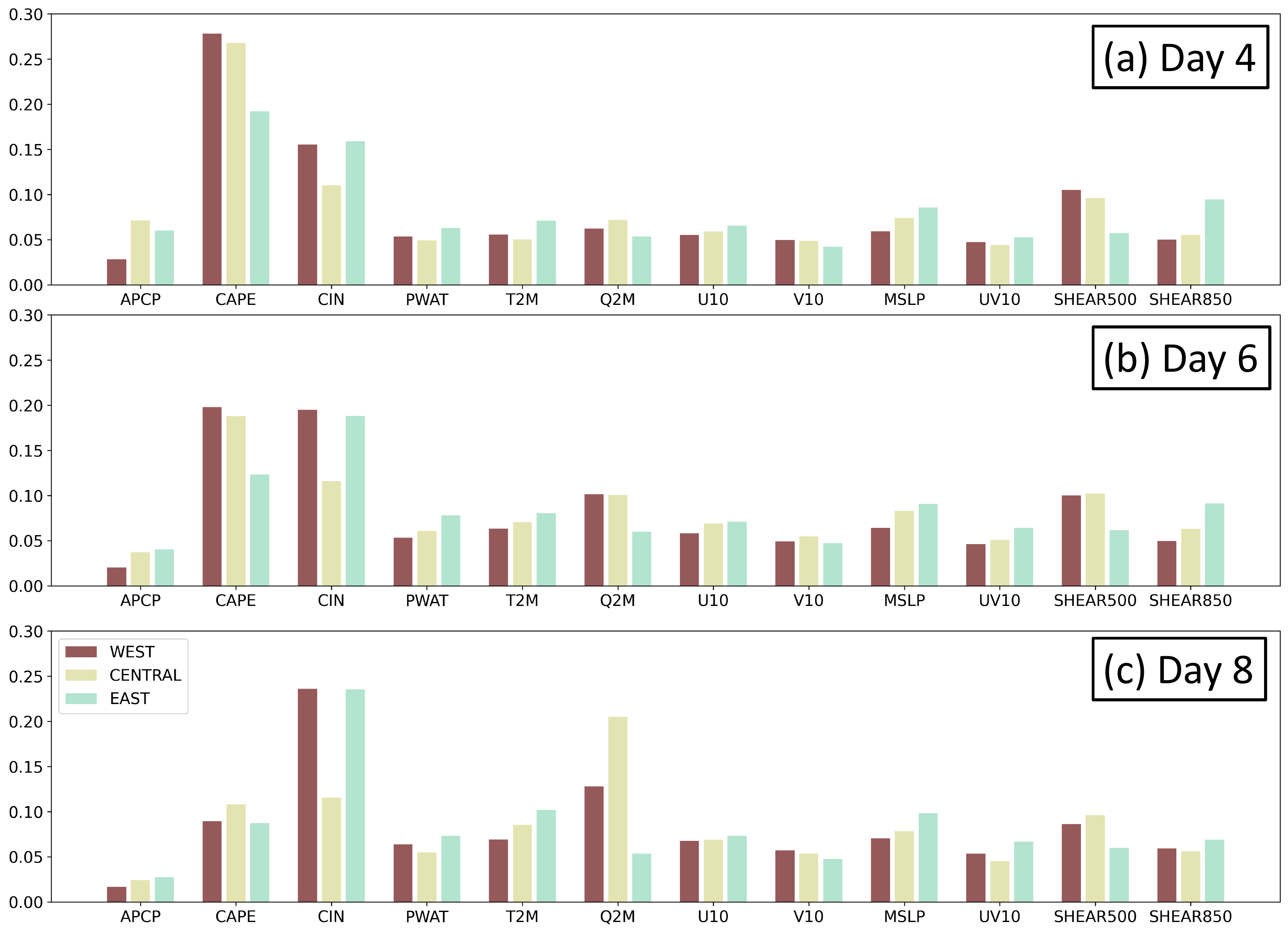}\\
 \caption{FIs for day (a)-(c) 4, 6, and 8 traditional CSU-MLP models grouped by meteorological variable and color coded by training model region as depicted in the legend.}\label{MR_GINI}
\end{figure*}

To better understand what the RFs have learned about severe weather prediction from the training process, FIs are aggregated by meteorological variable and region for the day 4, 6, and 8 CSU-MLP control models (Fig. \ref{MR_GINI}). Consistent with day 1-3 models developed by \citet{Hilletal2020}, CAPE, CIN, MSLP, SHEAR500, and SHEAR850 are the most importance predictors for the day 4 models (Fig. \ref{MR_GINI}a); CAPE is also less important in the East region as MSLP and SHEAR850 increases in importance, consistent with high-shear low-CAPE environments that are more prevalent in the southeast U.S. \citep{Sherburn2014hslc}. As lead time increases, CAPE becomes less important in the Central region (Fig. \ref{MR_GINI}b,c), being replaced by Q2M. Q2M also becomes more important in the West region, but CIN replaces CAPE as the most important predictor (e.g., Fig. \ref{MR_GINI}c). In the East, CIN also replaces CAPE as the most important predictor. 

\begin{figure*}[t]\centering
 \noindent\includegraphics[width=39pc,angle=0]{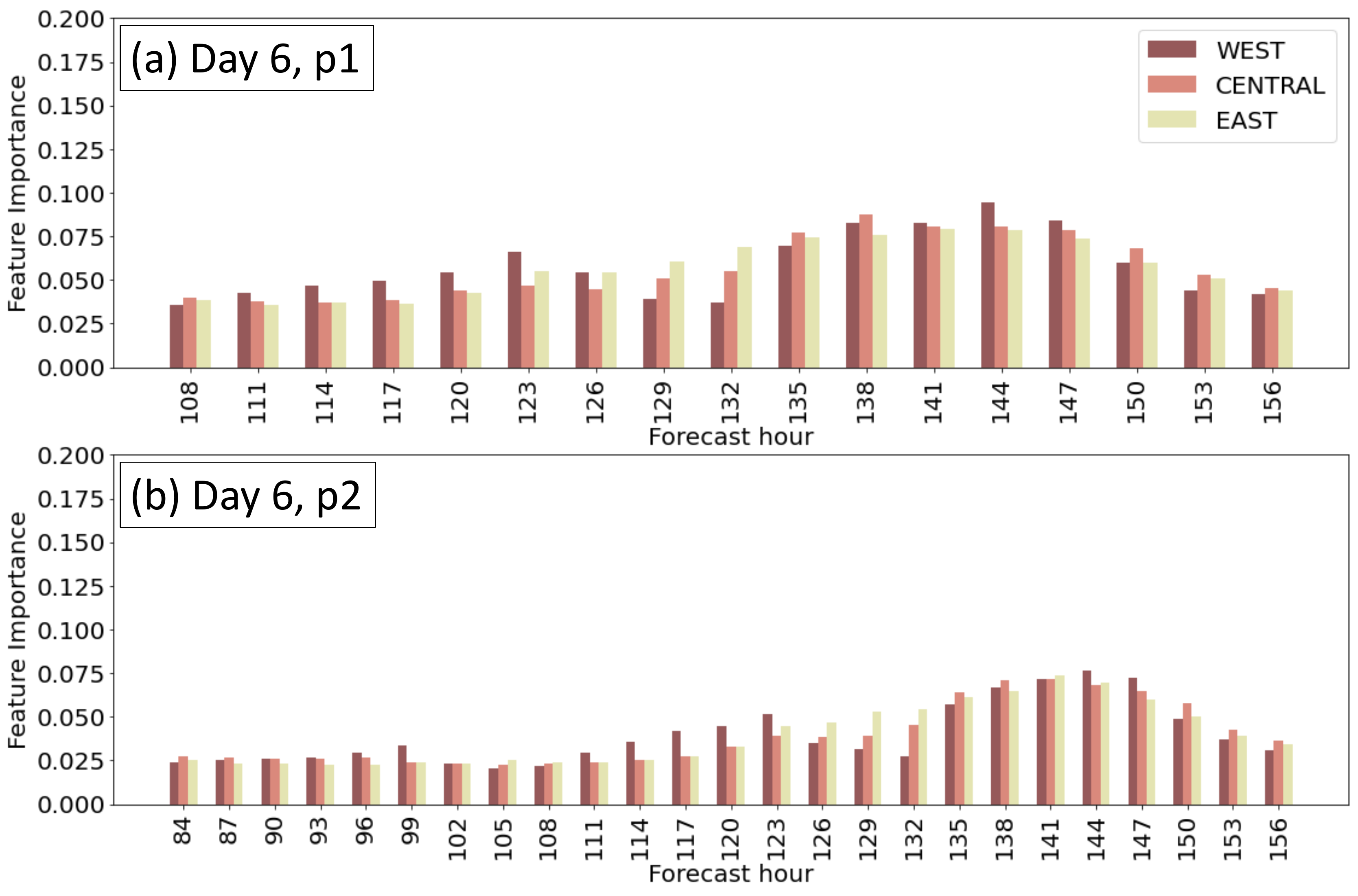}\\
 \caption{FIs for day 6 (a) p1 and (b) p2 models at each forecast hour and color coded by training region. FIs at each hour are aggregated across all meteorological variables.}\label{m1_m2}
\end{figure*}

FIs of the p1 and p2 models are explored to further understand how the RFs leverage features in the day(s) leading up to severe weather events. FIs for both models peak during the day 6 forecast window (i.e., forecast hours 132--156), but they also ramp up from the days leading up to the event (Fig. \ref{m1_m2}), suggesting the local meteorological environment is being used by the RFs in predictions. Slight differences in FIs exist by regional model as well. For example, the West region p1 and p2 models have secondary peaks near forecast hours 120--123 (Fig. \ref{m1_m2}a), approximately 00--03 UTC the day before the forecast window, and p2 has a tertiary peak near forecast hour 99 (Fig. \ref{m1_m2}b). Not only are the prior days variables being used, but there is a notable cyclical nature that matches the diurnal climatology of severe weather \citep[e.g.,][]{Hilletal2020}. In contrast, FIs in the Central and East regions do not have the same cyclic pattern, but rather have a nearly constant ramp up in FIs (e.g., orange and yellow bars in Fig. \ref{m1_m2}a). However, since these FIs are a summation over all meteorological predictors, it is not clear what aspects of the environment prior to a severe weather event are being learned by the RFs to make day 6 predictions. 

\begin{figure*}[t]\centering
 \noindent\includegraphics[width=39pc,angle=0]{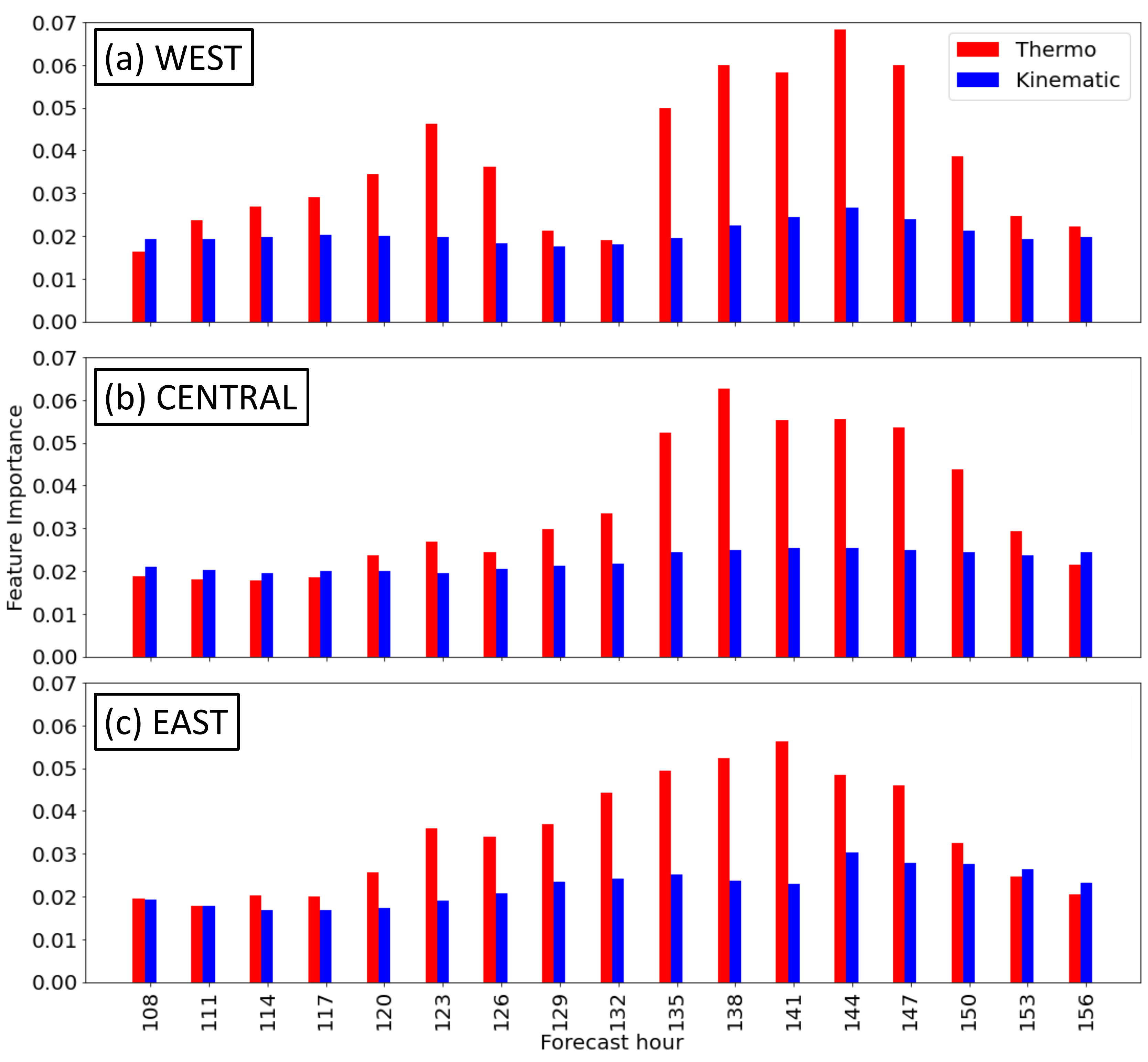}\\
 \caption{FIs for day 6 p1 models in the (a) WEST, (b) CENTRAL, and (c) EAST regions color coded by feature variable type as defined in Table \ref{t1}. Importances are summed at each forecast hour for the thermodynamica nd kinematic feature subsets.}\label{p1_REG}
\end{figure*}

To further clarify the day-prior FIs, features and FIs are separated into thermodynamic and kinematic subgroups (refer to Table \ref{t1}) for the p1 models (Fig. \ref{p1_REG}). In the West region models, which exhibited a strong cyclic FI pattern, the thermodynamic variables (e.g., CAPE, Q2M) are the primary contributor to the day-prior FI secondary peak, with a sharp increase at forecast hour 123 (Fig. \ref{p1_REG}a) -- the kinematic variable FIs in the West p1 model have a more subtle cyclic pattern, but still peak during the day 6 forecast window. In the Central region model, the FIs have a broad and uniform peak from forecast hours 135--147 and a smaller contribution compared to the West region in the day prior period (Fig. \ref{p1_REG}b). The East region models lean on day-prior thermodynamic predictors slightly more than the Central region models, with a longer ramp-up of FIs from forecast hours 123--141 (Fig. \ref{p1_REG}c), but the kinematic FIs for both the Central and East models are markedly smaller than the thermodynamic variable contributions. A full explanation for these FI patterns is reserved for future work, but as an initial assessment, the FIs highlight unique regional and meteorological relationships learned by the experimental models that were exploited by the RFs to make severe weather predictions. Changes to the local environment that preceeded severe weather events clearly influenced the RFs during training, but to what extent those variables contributed to forecast probabilities is not discernible from Gini importances alone. 

\section{Summary and Discussion}

Nine years of reforecasts from the GEFSv12 reforecast dataset are used along with historical records of severe weather to construct a novel RF prediction system capable of explictly and probabilistically predicting severe weather at 4--8 day lead times, i.e., the medium range. Human forecasts issued by the SPC are evaluated alongside the RF-based predictions to assess the operational utility of the ML forecasts. A handful of experiments are also conducted to explore whether forecasts could be improved through feature engineering and expanding the GEFSv12 ensemble size. The main conclusions are as follows:

\begin{enumerate}
    \item RF forecasts have more skill and higher resolution than the human-based outlooks, which is partly a reflection of the continuous probabilities of the RFs and their ability to issue lower probability contours more frequently that add considerable skill and resolution to the forecast system.
    \item The CSU-MLP forecasts tend to underforecast the occurrence of severe weather in the medium range at probabilistic contours above 15\% whereas SPC forecasts are calibrated prior to day 8. 
    \item Using spatially-averaged GEFS/R features yielded similarly skillful forecasts as the traditional CSU-MLP method while also allowing for prior-day meteorological information to inform the forecasts; the models learned to associate the buildup of thermodynamically- and kinematically-favorable environments with next-day severe weather events. Additionally, the similar skill amongst models suggests that spatiotemporally-displaced GEFS/R predictors are not particularly noisy but do not provide tremendous value to the RFs.
    \item Time-lagging the GEFSv12 reforecasts to produce larger initial ensembles for RF training degraded forecast skill but increased forecast resolution by generating larger areas of low probability forecasts.
    \item Feature importances revealed relationships known to be important for severe weather forecasting, providing confidence in the RF forecasts.
    \item The performance of the RF-based forecasts alongside the human-generated outlooks demonstrate their utility and potential value as a guidance tool in the human forecast process. 
\end{enumerate}

The comparisons between RF-based predictions and human forecasts provided in this work have some important caveats to consider. SPC forecasters have often employed specific philosophies in generating day 4--8 outlooks [Steven Weiss, personal communication] that likely limit the number of forecasts issued and hamper more skillful human-based medium-range forecasts. First, SPC forecasters are tasked with forecasting the probability of ``organized severe thunderstorms'', not necessarily severe weather reports, so they will never outline a high CAPE, low shear event in the medium range despite a high likelihood of thunderstorms. Second, SPC forecasters are very concerned with continuity and ramping up to an event. For example, forecasters may opt to introduce a forecast area in a day 3 or 2 outlook rather than days 4--8 because they may not want to highlight a threat area that has to be shifted, enlarged, or removed altogether in subsequent outlooks -- forecasters are hesitant to add outlook areas when confidence is too low, and they can add it in on the next forecast shift. Relatedly, SPC forecaster perception is that NWS Weather Forecast Offices do not like when SPC removes or reduces severe weather probabilities because it affects public messaging of severe weather. As a result, it is often common for SPC day 4--8 outlooks to be relatively small (e.g., Fig. \ref{COVERAGE}) and infrequent, particularly when atmospheric predictability for severe weather wanes in the warm season (e.g., Fig. \ref{MON_FREQ}). As confidence increases in a severe weather threat area, the probabilities can be increased and area expanded in day 1--3 outlooks. These and other internal constraints, along with the relative dearth of useful NWP model guidance, naturally restrict SPC forecast skill at longer lead times. On the other hand, the ML-guidance generated from the CSU-MLP could significantly aid SPC by increasing confidence and consistency to provide more lead time to operational partners and end users to the threat of severe weather.

The results presented highlight some ML success against baseline forecasts and also a number of unique avenues that could be explored moving forward to enhance and improve both ML-based guidance and the SPC human-based forecasts, as well as increase interpretability of the ML `black box'. While the feature assembly experiments (e.g., p2, tl10) did not yield forecasts that surpassed the skill of the traditional CSU-MLP system, the simplification of features could be exploited to include other ensemble diagnostic or summary metrics (e.g., mean, high or low member values) that characterize ensemble spread into the medium range. The meteorological predictors could also be varied in any of the ML configurations to explore which predictors add the most value, or objective methods (e.g., permutation importance) could be used to reduce feature redundancy and select a more optimal subset of features. It will also be vitally important that alternative interpretability metrics (e.g., tree interpreter \citep{Saabas2016}, Shapely additive values \citep[SHAP;][]{Shapley2016}, accumulated local effect \citep[ALE;][]{ApleyandZhu2020}) are employed to interrogate how the RFs make predictions; this exploration is underway and will be the focus of a follow-on manuscript. Additionally, the added benefit of the ML system against the underlying GEFS model could be quantified more explicitly. Traditionally, ML-based forecasts have been measured against the very model that generates the ML predictors, with demonstrated success improving upon the raw dynamical models \citep[e.g.,][]{herman2018money}. In this instance, with notable 2-m dry and low-instability biases in the GEFSv12 system\footnote{Internal SPC surveys have suggested these biases exist and are reducing forecaster confidence in deterministic Global Forecast System and GEFSv12 forecasts} \citep{Manikinetal2020}, it would be informative to quantify the value added by the ML system to correct for these biases when making severe weather predictions. 

Equipped with calibrated statistical products and expert human knowledge, SPC forecasters may be able to increase medium-range outlook skill by using the CSU-MLP RF forecasts and delineating lower-probability threat areas. Furthermore, by incorporating these types of robust, skillful statistical guidance products into the forecast process, SPC forecaster confidence in a forecast outcome may increase. Being able to unveil how and why the ML models are issuing probabilities in a certain area will provide additional confidence to forecasters to rely on the products as a forecast tool. We expect that the usefulness of the CSU-MLP prediction system and others like it is not necessarily limited to the medium-range either, and the applicability to subseasonal or seasonal predictions of severe weather is planned for future investigation. Additionally, with continued effort from the meteorological community to explain AI/ML methodologies (e.g., Chase et al. 2022) with comprehension, and to make these tools more common in academic settings, there will be more opportunities to pursue new and improved forecast methodologies. Finally, it is crucial that constant communication exists between ML developers and SPC forecasters to generate products that SPC operations finds useful and valuable. One such avenue includes continued participation and development of these ML products in the Hazardous Weather Testbed Spring Forecast Experiment \citep{Clark2021}.  

%%%%%%%%%%%%%%%%%%%%%%%%%%%%%%%%%%%%%%%%%%%%%%%%%%%%%%%%%%%%%%%%%%%%%
% TABLES---INSERT NEAR IN-TEXT DISCUSSION
%%%%%%%%%%%%%%%%%%%%%%%%%%%%%%%%%%%%%%%%%%%%%%%%%%%%%%%%%%%%%%%%%%%%%
%%  Enter tables near where they are discussed within the document. 
%%  Please place tables before/after paragraphs, not within a paragraph.
%%
%
%\begin{table}[t]
%\caption{This is a sample table caption and table layout.  Enter as many tables as
%  necessary at the end of your manuscript. Table from Lorenz (1963).}\label{t1}
%\begin{center}
%\begin{tabular}{ccccrrcrc}
%\hline\hline
%$N$ & $X$ & $Y$ & $Z$\\
%\hline
% 0000 & 0000 & 0010 & 0000 \\
% 0005 & 0004 & 0012 & 0000 \\
% 0010 & 0009 & 0020 & 0000 \\
% 0015 & 0016 & 0036 & 0002 \\
% 0020 & 0030 & 0066 & 0007 \\
% 0025 & 0054 & 0115 & 0024 \\
%\hline
%\end{tabular}
%\end{center}
%\end{table}

%%%%%%%%%%%%%%%%%%%%%%%%%%%%%%%%%%%%%%%%%%%%%%%%%%%%%%%%%%%%%%%%%%%%%
% FIGURES---INSERT NEAR IN-TEXT DISCUSSION
%%%%%%%%%%%%%%%%%%%%%%%%%%%%%%%%%%%%%%%%%%%%%%%%%%%%%%%%%%%%%%%%%%%%%
%%  Enter figures near where they are discussed within the document.
%%  Please place figures before/after paragraphs, not within a paragraph.
% %
%
%\begin{figure}[t]
%  \noindent\includegraphics[width=19pc,angle=0]{figure01.pdf}\\
%  \caption{Enter the caption for your figure here.  Repeat as
%  necessary for each of your figures. Figure from \protect\cite{Knutti2008}.}\label{f1}
%\end{figure}

\clearpage
%%%%%%%%%%%%%%%%%%%%%%%%%%%%%%%%%%%%%%%%%%%%%%%%%%%%%%%%%%%%%%%%%%%%%
% ACKNOWLEDGMENTS
%%%%%%%%%%%%%%%%%%%%%%%%%%%%%%%%%%%%%%%%%%%%%%%%%%%%%%%%%%%%%%%%%%%%%
\acknowledgments

This work is supported by the Joint Technology Transfer Initiative and NOAA Grant NA20OAR4590350. We would like to thank SPC forecasters for their invaluable perspectives about these forecast products and continued collaboration to develop cutting edge medium-range guidance products. 

%  Keep acknowledgments (note correct spelling: no ``e'' between the ``g'' and
% ``m'') as brief as possible. In general, acknowledge only direct help in
%  writing or research. Financial support (e.g., grant numbers) for the work done, 
%  for an author, or for the laboratory where the work was performed must be 
%  acknowledged here rather than as footnotes to the title or to an author's name.
%  Contribution numbers (if the work has been published by the author's institution 
%  or organization) should be placed in the acknowledgments rather than as 
%  footnotes to the title or to an author's name.

%%%%%%%%%%%%%%%%%%%%%%%%%%%%%%%%%%%%%%%%%%%%%%%%%%%%%%%%%%%%%%%%%%%%%
% DATA AVAILABILITY STATEMENT
%%%%%%%%%%%%%%%%%%%%%%%%%%%%%%%%%%%%%%%%%%%%%%%%%%%%%%%%%%%%%%%%%%%%%
% 
%
\datastatement

All RF-based forecasts are available upon request from the Colorado State University researchers, and will be made available in the near future in an online repository. SPC outlooks are available via a public archive at https://www.spc.noaa.gov/. The GEFS/R dataset and GEFSv12 forecasts are publicly available from Amazon AWS at https://registry.opendata.aws/noaa-gefs/.

%  The data availability statement is where authors should describe how the data underlying 
%  the findings within the article can be accessed and reused. Authors should attempt to 
%  provide unrestricted access to all data and materials underlying reported findings. 
%  If data access is restricted, authors must mention this in the statement. See
%  {http://www.ametsoc.org/PubsDataPolicy} for more info.

%%%%%%%%%%%%%%%%%%%%%%%%%%%%%%%%%%%%%%%%%%%%%%%%%%%%%%%%%%%%%%%%%%%%%
% APPENDIXES
%%%%%%%%%%%%%%%%%%%%%%%%%%%%%%%%%%%%%%%%%%%%%%%%%%%%%%%%%%%%%%%%%%%%%
%
%% If only one appendix, use

%\appendix

%% If more than one appendix, use \appendix[<letter>], e.g.,

%\appendix[A] 

%% Appendix title is necessary! For appendix title:

%\appendixtitle{Title of Appendix}

%%% Appendix section numbering (note, skip \section and begin with \subsection)
%
% \subsection{First primary heading}

% \subsubsection{First secondary heading}

% \paragraph{First tertiary heading}

%%%%%%%%%%%%%%%%%%%%%%%%%%%%%%%%%%%%%%%%%%%%%%%%%%%%%%%%%%%%%%%%%%%%%
% REFERENCES
%%%%%%%%%%%%%%%%%%%%%%%%%%%%%%%%%%%%%%%%%%%%%%%%%%%%%%%%%%%%%%%%%%%%%
% Make your BibTeX bibliography by using these commands:
\bibliographystyle{ametsocV6}
\bibliography{medium_range_forecasting}

\end{document}